\documentclass[pdftex,twocolumn,epjc3]{svjour3}          
\RequirePackage[T1]{fontenc}
\smartqed  
\RequirePackage{graphicx}
\RequirePackage{mathptmx}      
\RequirePackage{flushend}
\RequirePackage[numbers,sort&compress]{natbib}
\RequirePackage[colorlinks,citecolor=blue,urlcolor=blue,linkcolor=blue]{hyperref}
\RequirePackage{amsmath,amssymb}
\RequirePackage{multirow}
\RequirePackage{xspace}
\RequirePackage{xcolor}
\RequirePackage{cancel}
\RequirePackage[caption=false]{subfig}
\RequirePackage{subfig}
\RequirePackage{blindtext, subfig}
\RequirePackage{amsmath}
\usepackage{comment}

\allowdisplaybreaks

\journalname{Eur. Phys. J. C}

\newcommand{\ggf}{$gg$F\xspace}

\begin{document}

\title{The anomalous production of multi-leptons and its impact on the measurement of $Wh$ production at the LHC}

\author{Yesenia Hernandez\thanksref{e1,addr1}
            \and Mukesh Kumar\thanksref{e2,addr1}
             \and Alan S. Cornell\thanksref{e3,addr2}
             \and Salah-Eddine Dahbi\thanksref{e4,addr1}
             \and Yaquan Fang\thanksref{e5,addr3,addr4}
             \and Benjamin Lieberman\thanksref{e6,addr1}
             \and Bruce Mellado\thanksref{e7,addr1,addr5}
             \and Kgomotso Monnakgotla\thanksref{e8,addr1}
             \and Xifeng Ruan\thanksref{e9,addr1}
             \and Shuiting Xin\thanksref{e10,addr3,addr4}
}

\thankstext{e1}{e-mail: yesenia@cern.ch}
\thankstext{e2}{e-mail: mukesh.kumar@cern.ch}
\thankstext{e3}{e-mail: acornell@uj.ac.za}
\thankstext{e4}{e-mail: salah-eddine.dahbi@cern.ch}
\thankstext{e5}{e-mail: fangyq@ihep.ac.cn}
\thankstext{e6}{e-mail: benjamin.lieberman@cern.ch}
\thankstext{e7}{e-mail: bmellado@mail.cern.ch}
\thankstext{e8}{e-mail: jeremiah.kgomotso.monnakgotla@cern.ch}
\thankstext{e9}{e-mail: xifeng.ruan@cern.ch}
\thankstext{e10}{e-mail: shuiting.xin@cern.ch}

\institute{School of Physics and Institute for Collider Particle Physics, University of the Witwatersrand, Johannesburg, Wits 2050, South Africa.\label{addr1}
          \and
          Department of Physics, University of Johannesburg, PO Box 524, Auckland Park 2006, South Africa.\label{addr2}
          \and
           Institute of High Energy Physics, 19B, Yuquan Road, Shijing District, Beijing, China, 100049.\label{addr3}
           \and
           University of Chinese Academy of Sciences (CAS), 19A Yuquan Road, Shijing District, Beijing, China, 100049.\label{addr4}
           \and
           iThemba LABS, National Research Foundation, PO Box 722, Somerset West 7129, South Africa.\label{addr5}
}

\date{Received: date / Accepted: date}

\maketitle

\begin{abstract}
Anomalies in multi-lepton final states at the Large Hadron Collider (LHC) have been reported in 
Refs.~\cite{vonBuddenbrock:2017gvy,vonBuddenbrock:2019ajh}. These can be interpreted in terms of the production 
of a heavy boson, $H$,  decaying into a Standard Model (SM) Higgs boson, $h$, and a singlet scalar, $S$, which is 
treated as a SM Higgs-like boson. This process would naturally
affect the measurement of the $Wh$ signal strength at the LHC, where $h$ is produced in association with leptons and 
di-jets. Here, $h$ would be produced with lower transverse momentum, $p_{Th}$, compared to  SM processes. Corners of the phase-space are fixed according to the model parameters derived in Refs.~\cite{vonBuddenbrock:2016rmr,vonBuddenbrock:2017gvy} without additional tuning, thus nullifying potential look-else-where effects or selection biases.
Provided that no stringent requirements are made on $p_{Th}$ or related observables,  the signal strength of $Wh$ is $\mu(Wh)=2.41 \pm 0.37$. This corresponds to a deviation from the SM of $3.8\sigma$. This result further strengthens the need to measure with precision the SM Higgs boson couplings in $e^+e^-$, and $e^-p$ collisions, in addition to $pp$ collisions.
\end{abstract}

\section{Introduction}
\label{sec:intro}

The discovery of a Higgs boson ($h$)~\cite{Higgs:1964ia,Englert:1964et,Higgs:1964pj,Guralnik:1964eu} at the Large Hadron Collider (LHC) by the
ATLAS~\cite{Aad:2012tfa} and CMS~\cite{Chatrchyan:2012xdj} experiments has opened a new chapter in particle physics. Measurements provided 
so far indicate that the quantum numbers of this boson are consistent with those predicted by the Standard Model (SM)~\cite{Chatrchyan:2012jja,Aad:2013xqa}, 
and that the relative branching ratios (BRs) to SM particles follow what is predicted by the SM. With this in mind, a window of opportunity now opens for the 
search for new bosons and how these would affect the $h$ boson measurements. 

One of the implications of a  2HDM+S model, where $S$ is a scalar SM singlet, is the production of multiple-leptons through the decay chain $H\rightarrow Sh,SS$~\cite{vonBuddenbrock:2016rmr}, where $H$ is the heavy CP-even scalar and $h$ is the SM Higgs boson. Excesses in multi-lepton final states were reported in Ref.~\cite{vonBuddenbrock:2017gvy}. In order to further explore results with more data and new final states while avoiding biases and look-else-where effects, the parameters of the model were fixed in 2017 according to Refs.~\cite{vonBuddenbrock:2016rmr,vonBuddenbrock:2017gvy}. This includes setting the scalar masses as $m_H=270$\,GeV, $m_S=150$\,GeV,  treating $S$ as a SM Higgs-like scalar and assuming the dominance of the decays $H\rightarrow Sh,SS$. Excesses in opposite sign di-leptons, same-sign di-leptons, and three leptons, with and without the presence of $b$-tagged hadronic jets were reported in Ref.~\cite{vonBuddenbrock:2019ajh}.

In Ref.~\cite{Fang:2017tmh} the impact on the measurement of the process $H\to Sh$ was evaluated in final states including $h\to \gamma\gamma$ in association with 
hadronic jets. In particular, it was demonstrated that the impact on the measurement of $h$ produced via vector boson fusion (VBF) would be moderate, where the measurement 
of $h$ in association with $W\to jj$ would be affected significantly, as long as the transverse momentum of $h$, $p_{Th} < m_{W}$, where $m_{W}$ is the mass of the $W$ boson. 

In this article we expand on Ref.~\cite{Fang:2017tmh} by studying the potential impact on measurements related to $Wh, W\to jj/\ell \nu$ ($\ell=e,\mu$) and other relevant 
final states used in the measurement of the signal strength of $Wh$ by the LHC experiments. A survey of the existing measurements of the cross-section of the $Wh$ 
production mechanism from the ATLAS and CMS experiments is performed, with emphasis on measurements of the signal strength of the $Wh$ production mechanism 
in the corner of the phase-space where $p_{Th}<m_{W}$ is explored. Here we evaluate the size of the deviation from the SM in the production of $Wh$, as measured by 
the LHC experiments. The final states considered here were not included in the statistical analyses reported in Refs.~\cite{vonBuddenbrock:2017gvy,vonBuddenbrock:2019ajh}.

The paper is organised as follows: Section~\ref{sec:model} succinctly describes the simplified model used to model the BSM signal described above; Section~\ref{sec:method} 
reports on the available data and the methodology used to study it; Section~\ref{sec:inclusive} points to the compatibility of the results with the measurements of inclusive 
observables made by the experiments; Section~\ref{sec:results} summarises the findings of the paper and quantifies the size of the observed anomaly in the Higgs boson data. 

\section{The Simplified Model}
\label{sec:model}

In the model, the scalar $H$ has Yukawa couplings as it is assumed to be related to EW symmetry breaking (EWSB). 
The simplified Lagrangian used to describe the production of $H$ is:
\begin{align}
\mathcal{L}_{\text{H}} = -\frac{1}{4}~\beta_{g} \kappa_{_{hgg}}^{\text{SM}}~G_{\mu\nu}G^{\mu\nu}H
+\beta_{_V}\kappa_{_{hVV}}^{\text{SM}}~V_{\mu}V^{\mu}H. 
 \label{eqn:H_production}
\end{align}
These are the effective vertices required so that $H$ couples to gluons and the heavy vector bosons $V = W^\pm, Z$, respectively. The first term in~\ref{eqn:H_production} 
allows for the gluon fusion (\ggf) production mode of $H$, while the second term describes the VBF production mode of $H$  and $VH$ production mode. 
The $\kappa_{_{hgg}}^{\text{SM}}$ and $\kappa_{_{hVV}}^{\text{SM}}$ are the effective coefficients for the equivalent SM Higgs gluon fusion, and Higgs vector-boson fusion, 
whilst $\beta_g = y_{ttH}/y_{tth}$ is the scale factor with respect to the SM top-Yukawa coupling for $H$. Therefore, it is used for tuning the effective \ggf coupling. 
Similarly, $\beta_V$ represents the scale factor used to tune the $VVH$ couplings. 

On the other hand, the $S$ boson is assumed to only be produced through the $H$ decay so that its direct production is suppressed. The $S$ boson is included in this model 
as a singlet scalar that interacts with $H$ and the SM Higgs boson $h$. This allows the $H$ particle to produce $S$ bosons through the $H\to SS$ and $Sh$ decay modes. 
The assumption here considers the  $H\to Sh$ decay mode to have a 100\% BR. The effective interaction Lagrangians described in the following consider all these assumptions. 
The $S$ boson couples to the scalar sector as below:  
\begin{align}
{\cal L}_{HhS} = &-\frac{1}{2}~v\Big[\lambda_{_{hhS}} hhS + \lambda_{_{hSS}} hSS +
\lambda_{_{HHS}} HHS + \lambda_{_{HSS}} HSS \notag \\
&\qquad\quad+ \lambda_{_{HhS}} HhS\Big], 
\label{eqn:HS_coupling}
\end{align}
where the couplings are fixed to ensure that the BR for the $H\to Sh$ must satisfy the constraints discussed in~\cite{vonBuddenbrock:2018xar} . 
Furthermore, by fixing the parameters in the Lagrangian BRs of the Higgs-like $S$ boson are achieved. The effective interactions can be written as:
\begin{align}
{\cal L}_{S} =&\, \frac{1}{4} \kappa_{_{Sgg}} \frac{\alpha_{s}}{12 \pi v} S G^{a\mu\nu}G_{\mu\nu}^a
+ \frac{1}{4} \kappa_{_{S\gamma\gamma}} \frac{\alpha}{\pi v} S F^{\mu\nu}F_{\mu\nu} \notag \\
&+ \frac{1}{4} \kappa_{_{SZ\gamma}} \frac{\alpha}{\pi v} S Z^{\mu\nu}F_{\mu\nu}
+  \kappa_{_{SZZ}} m_Z S Z Z  \notag \\
&+  \kappa_{_{SWW}} m_W S W^{+} W^{-} - \sum_f \kappa_{_{Sf}} \frac{m_f}{v} S \bar f f. \label{eqn:S_decays}
\end{align}
Additionally, the couplings are globally re-scaled in order to suppress the direct production of $S$.

In the model the number of free parameters is reduced by fixing the BRs of $S$. For simplicity the BRs of $S$ are set to the same as that of a SM Higgs boson in the mass 
range considered here. In the above Lagrangian,  
$Z_{\mu\nu} = D_\mu Z_\nu - D_\nu Z_\mu$, $F_{\mu\nu}$ is the usual electromagnetic field strength tensor and $f$ refers to the SM fermions. 
Here, we neglect other possible terms for the self interaction of $S$ as they are not phenomenologically interesting for this study. 

It is also important to mention that the Lagrangians used here are the subset of full 2HDM+S models~\cite{vonBuddenbrock:2016rmr,vonBuddenbrock:2018xar, Muhlleitner:2016mzt}, where the 
couplings associated with particle spectrum of the model are functions of appropriate mixing angles of three CP-even scalars ($h, H, S$), a CP-odd scalar $(A)$ and 
charged scalar $(H^\pm)$. The parameters in Ref.~\cite{vonBuddenbrock:2018xar} also satisfy the: 
(a) theoretical constraints, like tree-level perturbative unitarity, the vacuum stability from global minimum conditions of the 2HDM+S potential and conditions which bound 
the potential from below; (b) the experimental constraints from $B\to X_s \gamma$ and $R_b$; and (c) the compatibility with the oblique parameters $S, T$ and $U$.  

\section{Methodology}
\label{sec:method}

The analyses for the associated production of $h$ with a $W$ or $Z$ bosons through Drell-Yan processes typically exploit the feature that $h$ is produced with larger 
transverse momentum than the SM background processes. The SM $Vh$ signal sensitivity is enhanced by considering corners of the phase-space with 
$p_{Th}>m_W$ where backgrounds can be strongly suppressed. This is actively used by the LHC experiments to effectively extract the $h$ signal for measurements of 
the signal strength. This implies that searches and measurements of $Wh$ at the LHC favor regions of the phase-space with $p_{Th}>m_W$ where a significantly large 
rate of $h$ can be produced. The high $p_{Th}$ restriction has to be taken into account if one is looking for deviations from the SM in the Higgs sector. This is achieved either 
by truncating the phase-space, excluding low $p_{Th}$ with large backgrounds, or by implementing multivariate analyses that include observables sensitive to $p_{Th}$, 
where the relative weight of large transverse momentum production is enhanced. 

By contrast, with the BSM signal $H\to S\,h$ with $m_H = 270$~GeV, $m_S = 150$~GeV and $m_h = 125$~GeV, $h$ displays significantly lower transverse momentum~\cite{vonBuddenbrock:2016rmr}. To a considerable degree, the $h$ signal produced via SM and BSM production mechanisms appears adjacent, 
but are distinct regions of the phase-space. The results provided by the ATLAS and CMS experiments pertain to the search and measurement of $Wh$ production in the SM 
and are not optimal for the search for new physics in general, and the BSM signal considered here, in particular. Nonetheless, a straw man approach is adopted here, whereby 
results that rely heavily on $p_{Th}$, or correlated observables, are discarded. Those results that explore the phase-space more ``inclusively" are considered here instead. 

It is important to reiterate that all considerations related to choice of phase-space or whether an analysis is discarded or not are based on a model with fixed parameters, as detailed in Ref~\cite{vonBuddenbrock:2016rmr,vonBuddenbrock:2017gvy} and dating back to 2017. This includes the above mentioned scalar  masses, securing the dominance of the $H\rightarrow Sh$ decay and considering $S$ as a SM Higgs-like scalar. This is a concerted effort in order not to scan of the phase-space, thus nullifying the potential biases or look-else-where effects.

Table~\ref{tab:AnalysesSummary} summarises the results from ATLAS and CMS experiments for the SM Higgs boson produced to date in association with a $W$ boson in 
leptonic  and di-jet final states. The reported signal strength ($\mu$) is provided by the respective publications. The Higgs decay modes considered here include $h\to WW$, $ZZ$, $\tau\tau$ and $\gamma\gamma$. Results from the $h\to b\bar{b}$ 
decay mode are not considered here as these analyses focus on large transverse momentum of the vector boson~\cite{VHbb_CMS,VHbb_ATLAS}. In the following the 
main event selection for each analysis is briefly described and the motivation for including the results in Section~\ref{sec:results} is discussed. The results included in the combination are selected by comparing the key kinematic distributions used in each analysis for the  $H\to Sh$ and SM $Wh$ 
processes from Monte Carlo simulation. Simulated events are generated with \texttt{PYTHIA}8~\cite{pythia8} using the NNPDF 2.3 LO~\cite{Ball:2012cx} for parton showering, 
with the A14 tune~\cite{ATL-PHYS-PUB-2014-021}, and without considering detector effects.

While the parameters of the model are fixed, we also present the kinematics of the final state with $m_H=250$\,GeV and $m_H=260$\,GeV, in addition to the nominal value.
The $H\to Sh$ samples are generated including $WW$, $ZZ$, $\tau\tau$ and $\gamma\gamma$ decay 
modes for the $S$ and $h$ bosons to  obtain the relevant final states with leptons, photons and jets for this study. Finally, the SM $Vh$ events are generated for each Higgs boson 
decay mode of interest separately. 
\begin{table*}[t]
\centering
\resizebox{1.78\columnwidth}{!}{
\renewcommand{\arraystretch}{1.5}
\begin{tabular}{lcccccccc} \hline
Higgs & \multirow{2}{*}{Ref.} & \multirow{ 2}{*}{Experiment} & $\sqrt{s}$,  \ensuremath{\mathcal{L}} &  Final & \multirow{ 2}{*}{Category} & \multirow{ 2}{*}{$\mu$}  & Used in & \multirow{ 2}{*}{Comments} \\  
decay & & & TeV,  fb$^{-1}$ & state & &  & combination & \\ \hline 
\multirow{12}{*}{$WW$} & \multirow{ 6}{*}{~\cite{ATLAS_HWW_Run1}} & \multirow{ 6}{*}{ATLAS} &  &  \multirow{3}{*}{2$\ell$} & DFOS  2j & $2.2^{+2.0}_{-1.9}$ & \checkmark ($Vh$) & \\ 
 & & & & & SS 1j  & $8.4^{+4.3}_{-3.8}$ & \checkmark ($Vh$) &  $2\ell$ combination: $\mu = 3.7^{+1.9}_{-1.5}$ \\ 
 & & & 7, ~4.5 & & SS 2j & $7.6^{+6.0}_{-5.4}$ & \checkmark ($Vh$)& \\ 
 & & & 8, 20.3 & \multirow{ 3}{*}{3$\ell$} & \multirow{2}{*}{1SFOS} & \multirow{ 2}{*}{$-2.9^{+2.7}_{-2.1}$} & \multirow{ 2}{*}{x} &  $m_{\ell_{0}\ell_{2}}$ used as input \\ 
 & & & & &  & & &   BDT discriminating variable\\
 & & & & & 0SFOS & $1.7^{+1.9}_{-1.4}$ & \checkmark ($Wh$) & \\  \cline{2-9}
& \multirow{ 2}{*}{~\cite{ATLAS_HWW_Run2}} & \multirow{ 2}{*}{ATLAS} & \multirow{ 2}{*}{13, 36.1}  & \multirow{ 2}{*}{3$\ell$} & 1SFOS & \multirow{ 2}{*}{$2.3^{+1.2}_{-1.0}$} & \multirow{ 2}{*}{\checkmark ($Wh$)} & 1SFOS channel uses $m_{\ell_{0}\ell_{2}}$ in the  \\
& & & & & 0SFOS & & & BDT but excess driven by 0SFOS \\  
\cline{2-9}
& \multirow{ 2}{*}{~\cite{CMS_HWW_Run1}} & \multirow{ 2}{*}{CMS} & 7, ~4.9 & 2$\ell$ & DFOS 2j & $0.39^{+1.97}_{-1.87}$ &  \checkmark ($Vh$) & Discrepancy at low $m_{\ell\ell}$ \\ 
& & & 8, 19.4 & 3$\ell$ & 0+1SFOS & $0.56^{+1.27}_{-0.95}$ &  \checkmark ($Wh$) &   \\  \cline{2-9}
& \multirow{ 2}{*}{~\cite{CMS_HWW_Run2}} & \multirow{ 2}{*}{CMS} & \multirow{ 2}{*}{13, 35.9} & $2\ell$ & DFOS 2j & $3.92^{+1.32}_{-1.17}$ &  \checkmark ($Vh$) & Discrepancy at low $m_{\ell\ell}$ \\ 
 & & & & 3$\ell$ & 0+1SFOS & $2.23^{+1.76}_{-1.53}$ &  \checkmark ($Wh$) &   \\ \hline\hline
\multirow{6}{*}{$\tau\tau$} & \multirow{ 2}{*}{~\cite{ATLAS_Htautau_Run1}} & \multirow{ 2}{*}{ATLAS} & \multirow{ 2}{*}{8, 20.3} & 1$\ell$ & $\ell + \tau_{\text{h}}\tau_{\text{h}}$ & $1.8\pm3.1$ & \checkmark ($Wh$) & \\  
 & & & & 2$\ell$ & $e^{\pm}\mu^{\pm}+ \tau_{\text{h}}$ & $1.3\pm2.8$ & \checkmark ($Wh$) &  \\ \cline{2-9}
& \multirow{ 2}{*}{\cite{CMS_Htautau_Run1}} & \multirow{ 2}{*}{CMS} & 7, ~4.9 & 1$\ell$ &  $\ell + \tau_{\text{h}}\tau_{\text{h}}$ & \multirow{ 2}{*}{$-0.33\pm1.02$} & x & BDT based on $p_{\text{T}}^{\tau_{\textrm{had}},{\textrm{lead}}}$ \\  
 & & & 8, 19.7 & 2$\ell$ & $e^{\pm}\mu^{\pm}+ \tau_{\text{h}}$  & & x & Split $p_{\text{T}}^{\ell_1} + p_{\text{T}}^{\ell_2} + p_{\text{T}}^{\tau}$ at 130~GeV \\ \cline{2-9}
& \multirow{ 2}{*}{\cite{CMS_Htautau_Run2}} & \multirow{ 2}{*}{CMS} & \multirow{ 2}{*}{13, 35.9} & 1$\ell$ & $\ell + \tau_{\text{h}}\tau_{\text{h}}$ & \multirow{ 2}{*}{$3.39^{+1.68}_{-1.54}$} & \multirow{ 2}{*}{\checkmark ($Wh$)} & \\ 
 & & & & 2$\ell$ & $e^{\pm}\mu^{\pm}+ \tau_{\text{h}}$ & & & \\ \hline\hline
\multirow{16}{*}{$\gamma\gamma$} & \multirow{ 3}{*}{~\cite{ATLAS_Hyy_Run1}} & \multirow{ 3}{*}{ATLAS} & \multirow{ 2}{*}{7, ~5.4} &  $\ell\nu$ & One-lepton & & &  \multirow{ 3}{*}{} \\ 
 & & & \multirow{2}{*}{8, 20.3} & $\cancel{\ell}\nu, \nu\nu$ & $E_{\textrm{T}}^{\textrm{miss}}$ &  $1.0\pm1.6$ & x & $E_{\textrm{T}}^{\textrm{miss}} > 70 - 100$~GeV \\ 
 & & &  & $jj$ & Hadronic & & & $p_{\textrm{Tt}}^{\gamma\gamma} > 70$~GeV \\ \cline{2-9}
& \multirow{ 3}{*}{~\cite{CMS_Hyy_Run1}} & \multirow{ 3}{*}{CMS} & \multirow{ 2}{*}{7, ~5.1} &  $\ell\nu$ & One-lepton & & & Split  $E_{\textrm{T}}^{\textrm{miss}}$ at $45$~GeV \\ 
 & & & \multirow{2}{*}{8, 19.7}& $\cancel{\ell}\nu, \nu\nu$ & $E_{\textrm{T}}^{\textrm{miss}}$ &  $-0.16^{+1.16}_{-0.79}$ & x & $E_{\textrm{T}}^{\textrm{miss}} > 70$~GeV\\ 
 & & & & $jj$ & Hadronic & & & $p_{\textrm{T}}^{\gamma\gamma} > 13m_{\gamma\gamma}/12$ \\ \cline{2-9}
 & \multirow{ 4}{*}{~\cite{ATLAS_Hyy_Run2_New}} & \multirow{ 4}{*}{ATLAS} & \multirow{ 4}{*}{13, 139} &  \multirow{ 2}{*}{$\ell\nu$} & \multirow{ 2}{*}{One-lepton} &  $2.41^{+0.71}_{-0.70}$  &   \checkmark ($Wh$) & $p_{\textrm{T}}^{\ell + E_{\textrm{T}}^{\textrm{miss}}} < 150$~GeV \\ 
 & & & & & & $2.64^{+1.16}_{-0.99}$ & x & $p_{\textrm{T}}^{\ell + E_{\textrm{T}}^{\textrm{miss}}} > 150$~GeV \\ 
 & & &  & \multirow{ 2}{*}{$jj$} & \multirow{ 2}{*}{Hadronic} & $0.76^{+0.95}_{-0.83}$ & x & $60<m_{jj}<120$~GeV \\
 & & & & & & $3.16^{+1.84}_{-1.72}$ & \checkmark (VBF+$Vh$) &  $m_{jj}\, \in \,[0,\,60]\,||\,[120,\,350]$~GeV \\  \cline{2-9} 
& \multirow{ 3}{*}{~\cite{CMS_Hyy_Run2}} & \multirow{ 3}{*}{CMS} & \multirow{ 3}{*}{13, 35.6} &  \multirow{ 1}{*}{$\ell\nu$} & \multirow{ 1}{*}{One-lepton} &  $3.0^{+1.5}_{-1.3}$  & \multirow{ 1}{*}{x} & \multirow{1}{*}{Superseeded by full Run 2 result}\\ 
 & & &  & $\cancel{\ell}\nu, \nu\nu$ & $E_{\textrm{T}}^{\textrm{miss}}$ & - & x & $E_{\textrm{T}}^{\textrm{miss}} > 85$~GeV \\
 & & & & $jj$ & Hadronic & $5.1^{+2.5}_{-2.3}$ & \checkmark ($Vh$) &  $p_{\textrm{T}}^{\gamma\gamma}/m_{\gamma\gamma}$ not used \\
 \cline{2-9}
 & \multirow{ 2}{*}{~\cite{CMS_Hyy_Run2_FullRun2}} & \multirow{ 2}{*}{CMS} & \multirow{ 2}{*}{13, 137} &  \multirow{ 1}{*}{$\ell\nu$} & \multirow{ 1}{*}{One-lepton} &  $1.31^{+1.42}_{-1.12}$  & \multirow{ 1}{*}{\checkmark ($Wh$)} & \multirow{1}{*}{$p_\textrm{T}^{V} < 75$~GeV}\\ 
 & & & & $jj$ & Hadronic & $0.89^{+0.89}_{-0.91}$ & x &  $p_{\textrm{T}}^{\gamma\gamma}/m_{\gamma\gamma}$ used in BDT \\
 \hline\hline
 \multirow{5}{*}{ZZ} &  \multirow{2}{*}{~\cite{ATLAS_ZZ_Run2}} &  \multirow{2}{*}{ATLAS} &  \multirow{2}{*}{13, 139} & $\ell\ell\ell\ell + \ell \nu$ &  Lep-enriched & \multirow{2}{*}{$1.44^{+1.17}_{-0.93}$} &  \multirow{2}{*}{x} &  Number of jets used in MVA \\
 & & & & $\ell\ell\ell\ell + q\bar{q}$ & $2j$ & &  & $m_{jj}$ used in MVA\\ 
 \cline{2-9} 
  &  \multirow{3}{*}{~\cite{CMS_ZZ_Run2}} &  \multirow{3}{*}{CMS} &  \multirow{3}{*}{13, 137.1} & \multirow{2}{*}{$\ell\ell\ell\ell + \ell \nu$} &  Lep-low $p_{\textrm{T}}^{h}$ & $3.21^{+2.49}_{-1.85}$ &  \checkmark ($Vh$) &  $p_{\textrm{T}}^{h}<$150~GeV \\
  & & & & & Lep-high $p_{\textrm{T}}^{h}$ & $0.00^{+1.57}_{-0.00}$ & x & $p_{\textrm{T}}^{h}>$150~GeV \\
 & & & & $\ell\ell\ell\ell + q\bar{q}$ & $2j$ & $0.57^{+1.20}_{-0.57}$ & x & $60<m_{jj}<120$~GeV \\ \hline
\end{tabular}}
\caption{\label{tab:AnalysesSummary} Summary of ATLAS and CMS $Vh$ results. The ``-" symbol indicates that the signal strength result is not provided for that specific category.}
\end{table*}

\subsection{$Vh\to VWW$}
\label{sec:ww}
The $Wh$ results in the $h\to WW^{\ast}$ decay using the Run 1 data sample collected at the ATLAS detector are obtained in two- and three-lepton final 
states~\cite{ATLAS_HWW_Run1}, denoted in the following as $2\ell$ and $3\ell$, respectively. The former requires exactly two well isolated leptons with high 
transverse momentum and is further split in different-flavour opposite-sign (DFOS) and same-sign (SS) $2\ell$ channels. 

In the DFOS $2\ell$ category the vector boson (either a $W$ or $Z$ boson) associated to the Higgs boson decays hadronically and produces two jets, while 
the $e^{\pm}\mu^{\mp}$ pair originates from the $h\to WW^{\ast}$ process. The SS $2\ell$ channel targets $Wh$ production when the $W$ boson that radiates 
the Higgs boson decays leptonically, while one of the $W$ bosons coming from $h\to WW^{\ast}$ decays hadronically, and the other - with same charge as the 
former $W$ boson - decays leptonically. In both categories lower bounds on the invariant mass of the lepton pair ($m_{\ell\ell}$) and on the missing transverse 
energy ($E_{\textrm{T}}^{\textrm{miss}}$) are applied, as well as a veto on events with the presence of $b$-tagged jets. For DFOS $2\ell$ events, several constraints 
on the dijet kinematics are required to select jets associated to $W/Z$ bosons. The rapidity separation between the two highest $p_{\textrm{T}}$ jets, $\Delta y_{jj} < 1.2$, 
and the invariant mass of these two jets, $|m_{jj} - 85| < 15$~GeV, are imposed. Finally, the selection exploits the kinematics of the lepton pair to be consistent with the 
$h\to WW^{\ast}$ decay, so the  azimuthal angular separation between the two leptons ($\Delta \phi_{\ell\ell}$) is required to be below 1.8~rad and $m_{\ell\ell} < 50$~GeV. 
In the SS $2\ell$ channel a further categorisation divides the events by having exactly one jet or exactly two jets in the final state.  Similarly to the DFOS category, a set of 
requirements on the minimum invariant mass of a lepton and a jet, the smallest opening angle between the lepton which minimises the above variable and a jet, and the 
transverse mass of the leading lepton and the $E_{\textrm{T}}^{\textrm{miss}}$ ($m_{\textrm{T}}$) are used. All these channels present an observed signal strength which is 
above the unity by one to two standard deviations ($\sigma$), as observed in Table~\ref{tab:AnalysesSummary}. The measured signal strength of the 2$\ell$ categories in 
ATLAS using Run 1 data results in $3.7^{+1.9}_{-1.5}$~\cite{ATLAS_HWW_Run1}. 
This result will be used in this paper.   

\begin{figure}[t]
\centering
\subfloat[]{\includegraphics[width=0.45\textwidth]{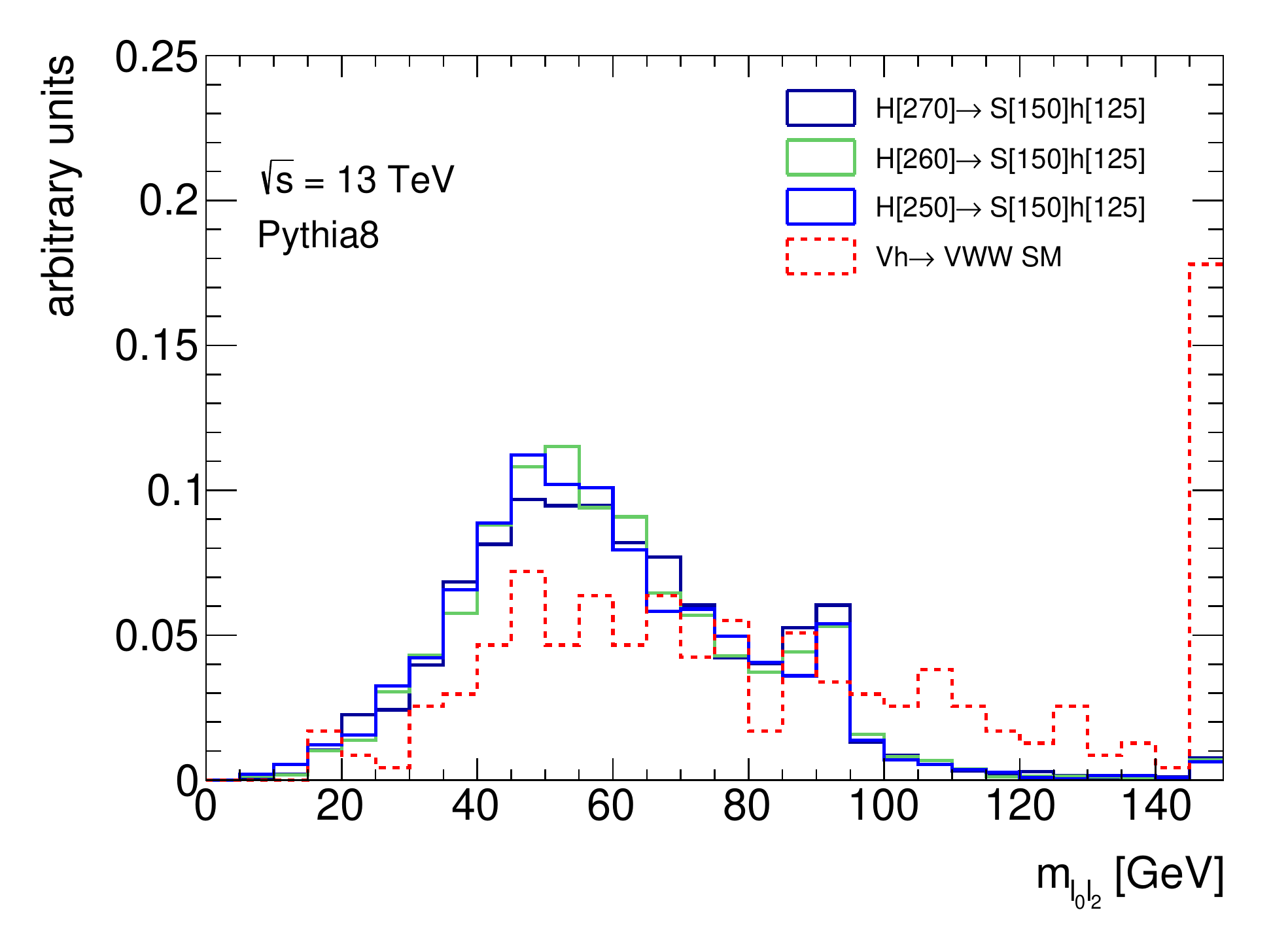}
\label{fig:Mll02}}\\
\subfloat[]{\includegraphics[width=0.45\textwidth]{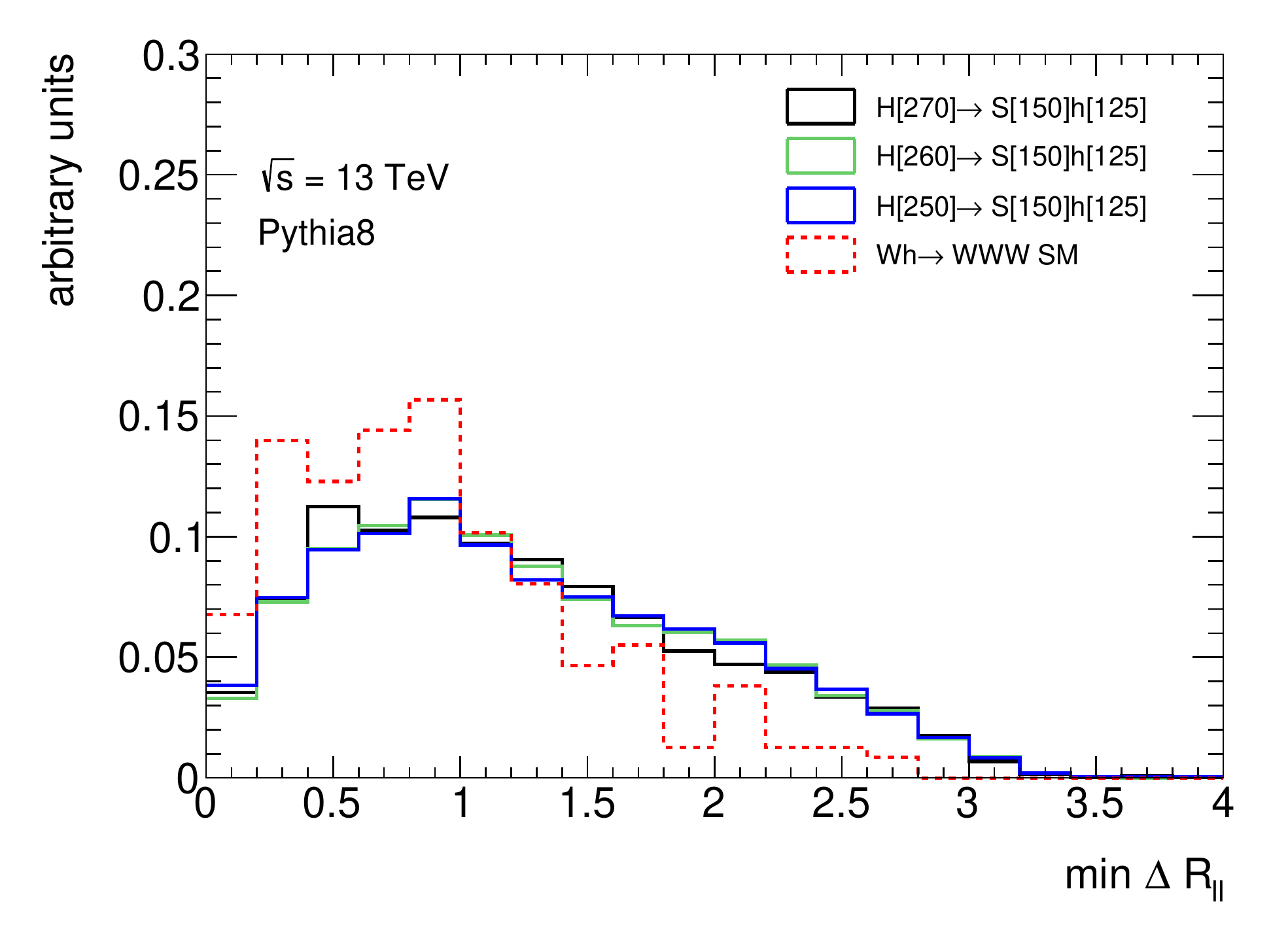}
\label{fig:minDRll}}
\caption{Invariant mass of the lepton with different electric charge and the lepton originating from the $W$ boson radiating the SM Higgs particle (a) and minimal distance between leptons (b) in the $Wh$ $3\ell$ 
channel for several $H\to Sh$ samples (solid lines) compared with the SM $Vh$ process with $h\to WW$ (dashed line) generated with \texttt{PYTHIA8}. 
The last bin contains overflow events. }
\end{figure}

In the $3\ell$ channel the $W$ bosons are expected to decay leptonically. These events are selected by having exactly three leptons with total charge of $\pm 1$ and at most 
one jet in the final state. Events are further categorised depending on the presence of same-flavour opposite-sign (SFOS) lepton pairs: 0SFOS and 1SFOS. The 0SFOS category 
includes $e^{\pm}e^{\pm}\mu^{\mp}$ and $\mu^{\pm}\mu^{\pm}e^{\mp}$ final states. These types of events highly benefit from low background contamination and no additional 
selection is applied. The angular separation of the Higgs decay lepton candidates ($\Delta R_{\ell\ell}$) is used in the likelihood fit to extract the results. The observed signal strength 
of the 0SFOS category is $1.7^{+1.9}_{-1.4}$ and it will be considered in the results section. Events with at least 1SFOS lepton pair require $\Delta R_{\ell\ell} < 2$ and the invariant 
mass of all SFOS combinations must satisfy $|m_{\ell^{\pm}\ell^{\mp}} - m_Z| > 25$~GeV in order to reject $WZ$ and $ZZ$ events. In addition, a multivariate discriminant based on 
Boosted Decision Trees (BDT)~\cite{hoecker2007tmva,pedregosa2012scikitlearn} is used. An important BDT input discriminating variable is the invariant mass of the lepton with 
different electric charge and the lepton originated from the $W$ boson radiating the SM Higgs particle ($m_{\ell_{0}\ell_{2}}$). This quantity tends to lower values for $H\to Sh$ events 
compared with the $Wh$ process as shown in Figure~\ref{fig:Mll02} for events with exactly three leptons with $p_{\textrm{T}} > 25,20,15$~GeV and total electric charge of $\pm 1$. The same behavior is also observed for $WZ^{\ast}$ and $Z$+jets events. These are the dominant background 
contributions for this category and they are mostly located in the $m_{\ell_{0}\ell_{2}} < 100$~GeV region. Given this feature, it is expected that the BDT discriminates these SM 
background processes, as well as the $H \to Sh$ signal, to the benefit of the target decay: $Wh \to WWW$. In light of this, the observed signal strength in 1SFOS events will not be 
combined with results from other categories. 

ATLAS has also published more recent $Wh$ results using 36.1~fb$ ^{-1}$ from the Run 2 dataset~\cite{ATLAS_HWW_Run2} for which only 3$\ell$ channels are considered. 
The selection strategy follows that from Run 1, but the usage of multivariate techniques has also been extended to the 0SFOS channel. In this case two BDTs are developed to 
reject $WZ$ and $t\bar{t}$ events. Mostly leptonic kinematic variables are used as inputs to the  BDT against $WZ$ backgrounds in the 0SFOS category from which only three 
are common to the 1SFOS category: the invariant mass of the Higgs lepton candidates, $E_{\textrm{T}}^{\textrm{miss}}$ and the difference in pseudo-rapidity between the leptons with the same electric charge. The BDT against $t\bar{t}$ uses as input variables hadronic quantities such as the number of jets and the transverse momentum of the jet with
highest $p_{\textrm{T}}$. The observed signal strength combining all $3\ell$ channels shows a deviation of about 2$\sigma$ with respect to the SM expectation, as quoted in
Table~\ref{tab:AnalysesSummary}, and it will be used in the combination in Section~\ref{sec:results}. Although the channel with at least 1SFOS lepton pair still makes use of the $m_{\ell_{0}\ell_{2}}$ as 
the BDT input discriminating variable, it can not be isolated and excluded from the $Vh$ combination exercise. It is important to note that the 0SFOS category alone would 
provide a higher discrepancy, as in this case the $H\to Sh$ is not expected to be rejected by the selection criteria. 
However, the observed  signal strength result from Ref.~\cite{ATLAS_HWW_Run2} combines both categories so the result for 0SFOS events can not be accounted for separately.

\begin{figure}[t]
\centering\includegraphics[width=0.45\textwidth]{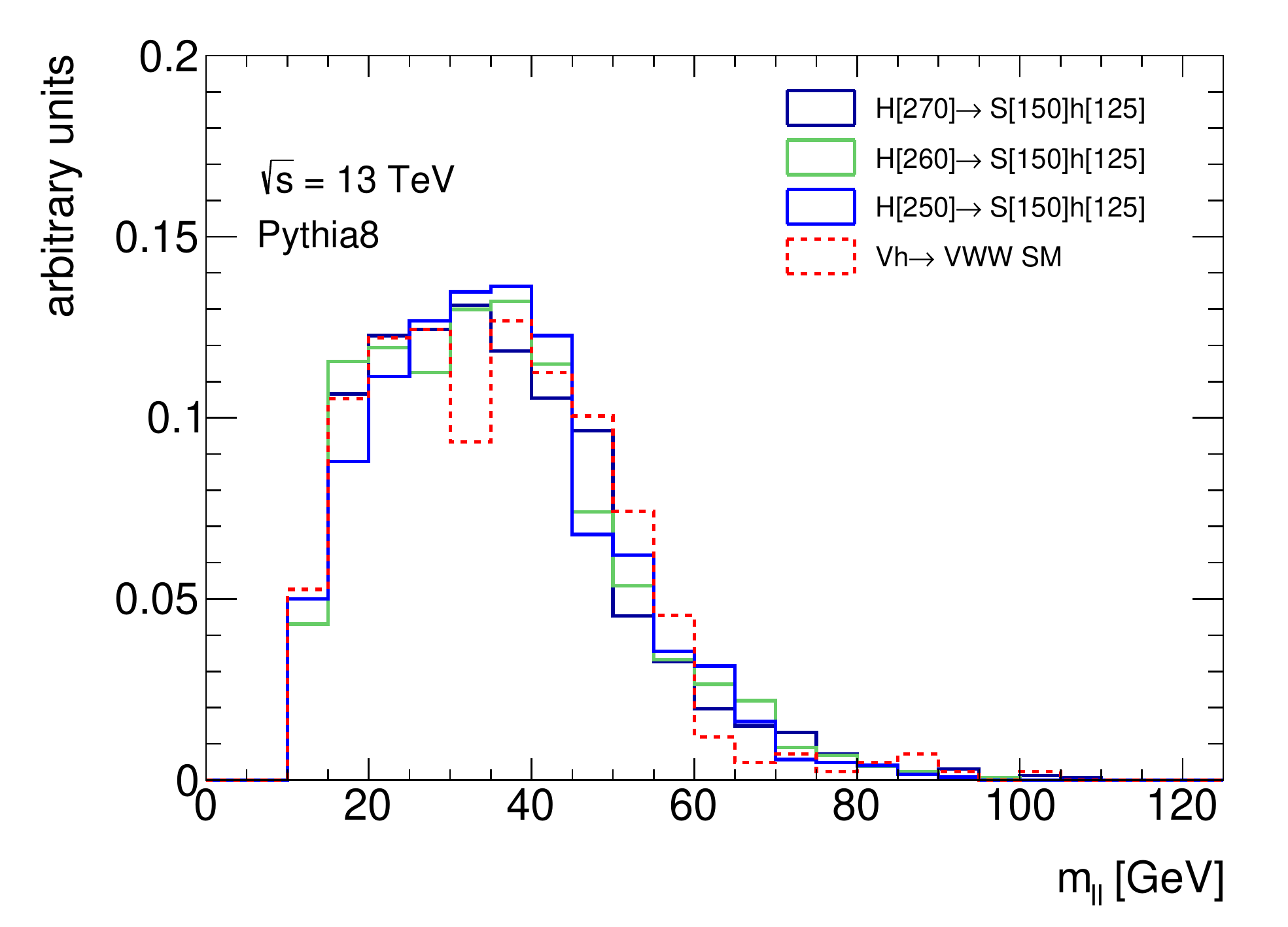}
\caption{Di-lepton invariant mass for several $H\to Sh$ samples (solid lines) compared with the SM $Vh$ process with $h\to WW$ (dashed line) 
generated with \texttt{PYTHIA8}.}
\label{fig:Mll}
\end{figure}
 
The CMS collaboration has also published results for the $Vh$ production mode with $h\to WW^{\ast}$ decay using Run 1 and partial Run 2 
datasets~\cite{CMS_HWW_Run1,CMS_HWW_Run2}. In these results a $Vh$ tagged category is defined by selecting events with a DFOS lepton pair with at least two 
jets in the final state. Similar to the ATLAS Run 1 strategy, $m_{jj}$ is used to guarantee the consistency with the parent boson mass and $|\Delta \eta_{jj}|<3.5$ is applied to 
avoid overlap with VBF events. In addition, the leptons are required to have small $\Delta R_{\ell\ell} $ since they are expected to be emitted in nearby directions due to the 
spin-0 nature of the SM Higgs boson. Finally, $m_T$ is required to be between 60~GeV and the mass of the SM Higgs boson. The $m_{\ell\ell}$ distribution is used as an input 
for the template fit to obtain the signal strength results. Both Run 1 and Run 2 results show a discrepancy between the observed data and the SM expectation 
at $m_{\ell\ell} < 50$~GeV. The SM Higgs boson as well as the $H\to Sh$ process are both expected to concentrate at the low $m_{\ell\ell}$ region as shown in Figure~\ref{fig:Mll}. 
As quoted in Table~\ref{tab:AnalysesSummary}, the signal strength is below unity for the Run 1 analysis, while the observed Run 2 data  presents an excess of $\sim$2.2$\sigma$. 
Since the selection is the same in both cases there is no reason to select one result and reject the other. In light of the CMS event selection, the observed signal strengths from the 
DFOS category using Run 1 and Run 2 datasets will both be used in the combination. 

Finally, CMS also targets events in the $3\ell$ category which are further split into two subcategories based on the existence of SFOS lepton pairs in the triplet. Opposite to 
ATLAS, the use of multivariate techniques is not considered by the CMS strategy. 
To reduce 
Drell-Yan processes a lower bound on the $E_{\textrm{T}}^{\textrm{miss}}$ and a $Z$ boson veto are applied for 1SFOS events. The observed signal strength for this category 
is extracted using the minimum $\Delta R_{\ell\ell}$ between oppositely charged leptons in the likelihood fit (see Figure~\ref{fig:minDRll}). Table~\ref{tab:AnalysesSummary} shows the same trend as previously discussed for the $2\ell$ channel: Run 1 results 
present a signal strength below one but fully consistent with the SM due to the large uncertainty. The situation is the opposite with the partial Run 2 dataset for which the signal 
strength is above unity, with a deviation from the SM expectation of $\sim$1.3$\sigma$. As discussed for the $2\ell$ category, both Run 1 and Run 2 results from CMS will be 
included in the combination. 

\subsection{$Wh\to W\tau\tau$}
\label{sec:tata}
Results for the associated production of the SM Higgs boson with a $W$ boson, where the Higgs boson is decaying into a pair of tau leptons have been performed by the 
ATLAS and CMS collaborations. The strategy in both experiments split the events into two categories, depending on the number of tau leptons decaying to hadrons ($\tau_{\textrm{had}}$), 
while the $W$ boson is assumed to decay leptonically. In the first category, the selection requires one electron and one muon with the same electric charge; and the presence of one
$\tau_{\textrm{had}}$ candidate in the final state ($e^{\pm}\mu^{\pm}\tau_{\textrm{had}}$). The second category selects events having one electron or muon accompanied by two 
$\tau_{\textrm{had}}$ candidates from the SM Higgs decay ($\ell\tau_{\textrm{had}}\tau_{\textrm{had}}$). 

\begin{figure}[t]
\centering
\subfloat[]{\includegraphics[width=0.45\textwidth]{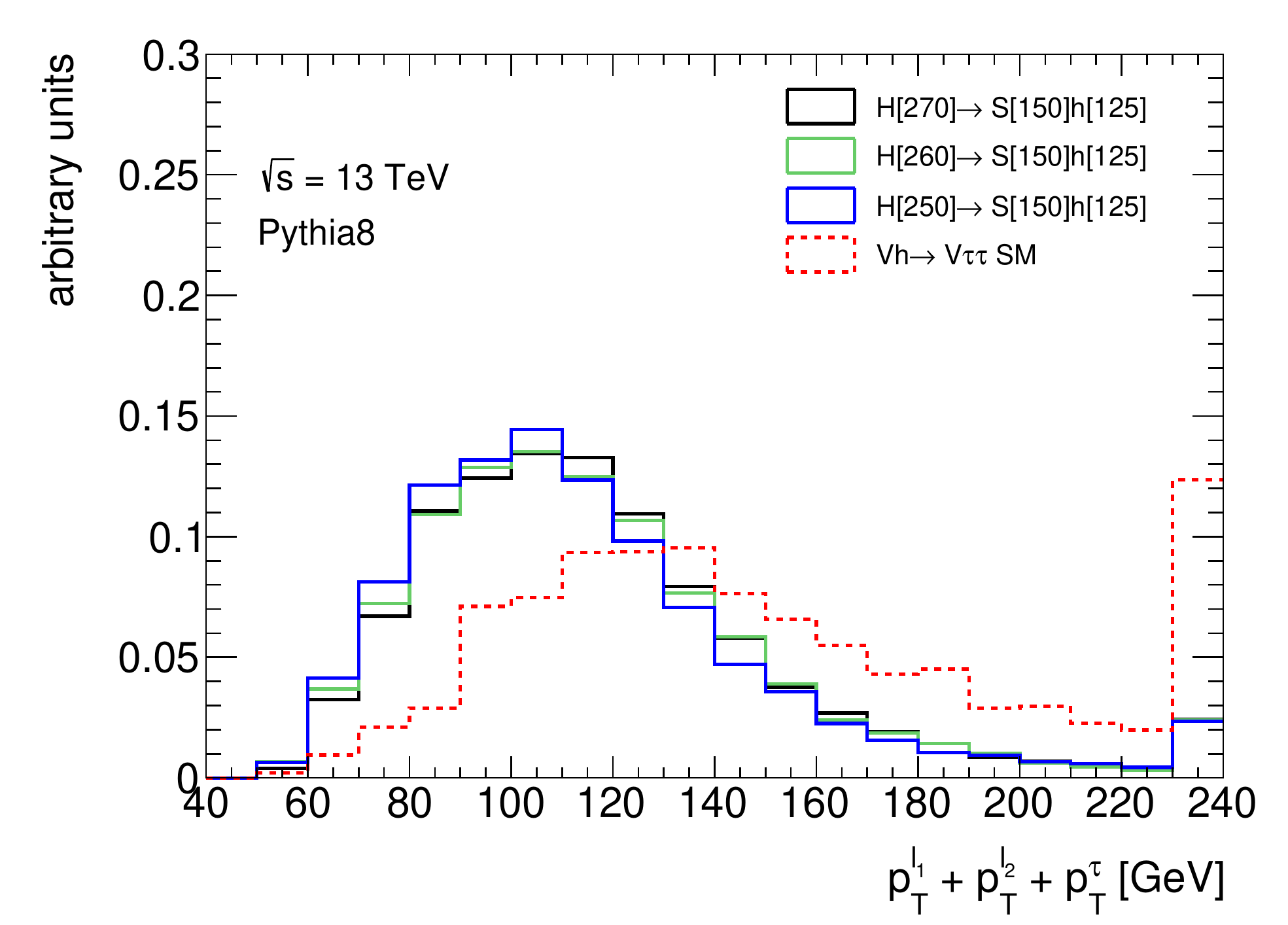}
\label{fig:LT_OneHadTau}}\\
\subfloat[]{\includegraphics[width=0.45\textwidth]{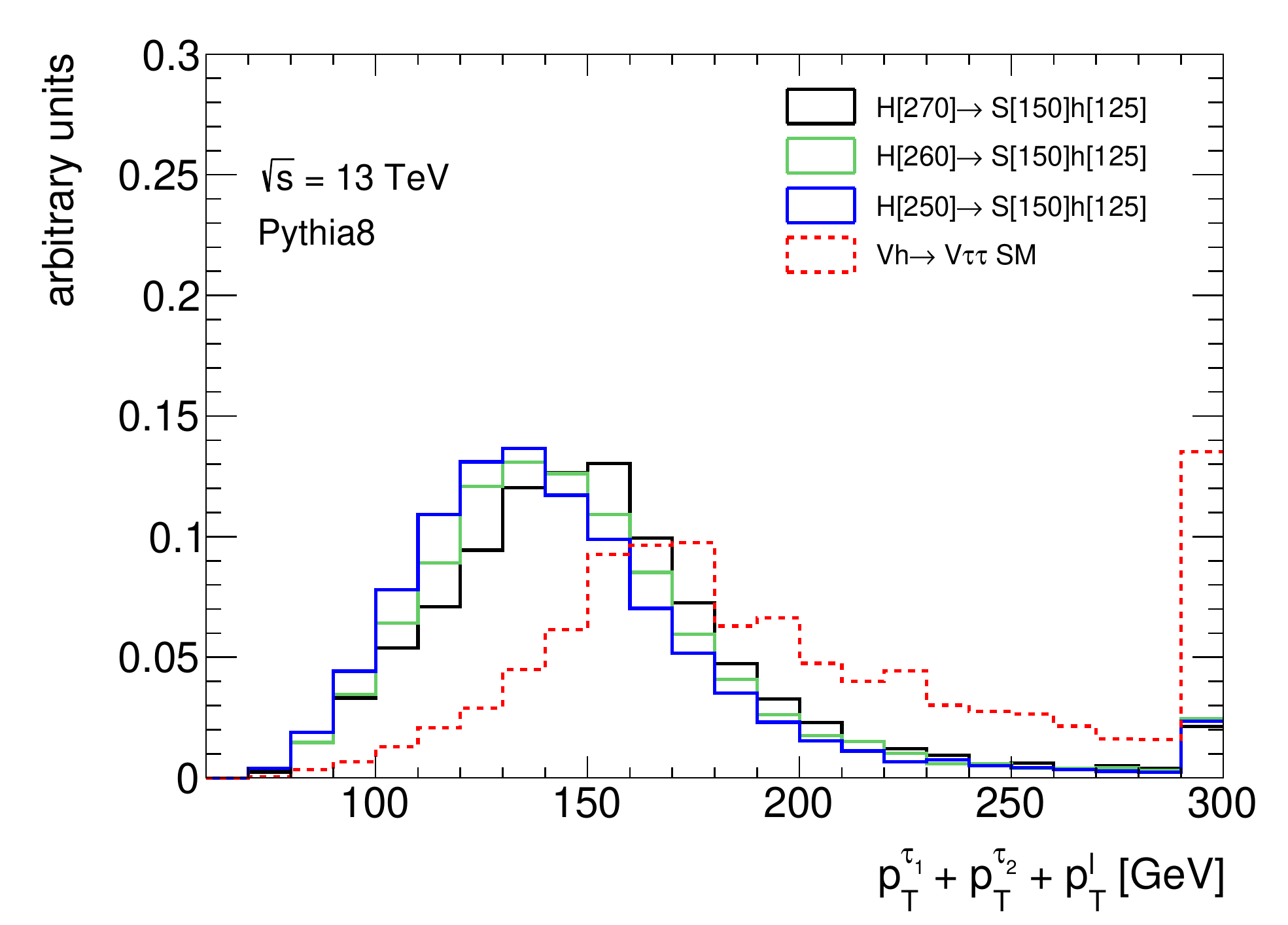}
\label{fig:LT_TwoHadTau}}
\caption{Scalar sum of the transverse momentum of two leptons and a hadronic tau (a) and scalar sum of the of transverse momentum of the lepton 
and two hadronic taus (b) for several $H\to Sh$ samples (solid lines) compared with the SM $Vh$ process with $h\to \tau\tau$ (dashed line) generated 
with \texttt{PYTHIA8}. The last bin contains overflow events.}
\end{figure}

The results from ATLAS are obtained using the Run 1 dataset~\cite{ATLAS_Htautau_Run1}. The kinematic selection for the $e^{\pm}\mu^{\pm}\tau_{\textrm{had}}$ 
category requires the scalar sum of the $p_{\textrm{T}}$ of the electron, muon and $\tau_{\textrm{had}}$ to be greather than 80~GeV.
Figure~\ref{fig:LT_OneHadTau} shows the scalar sum of the leptons' $p_{\textrm{T}}$ for events with exactly one electron and one muon satisfying $p_{\textrm{T}}^{\ell} > 20, 10$~GeV; and one hadronic tau with $p_{\textrm{T}}^{\tau} > 20$~GeV. 
It is clear that the lower bound threshold on this quantity keeps most of the $Wh$ and $H\to Sh$ processes. 
In the $\ell\tau_{\textrm{had}}\tau_{\textrm{had}}$ category the transverse mass of the lepton 
and $E_{\textrm{T}}^{\textrm{miss}}$ is required to be above 20~GeV and the two $\tau_{\textrm{had}}$ candidates must be within a $\Delta R$ of 2.8. Finally, the scalar sum of 
the $p_{\textrm{T}}$ of the lepton and the two $\tau_{\textrm{had}}$ is required to be above 100~GeV. Figure~\ref{fig:LT_TwoHadTau} compares the spectrum of this variable for 
events with one electron or muon with $p_{\textrm{T}}^{\ell} > 24$~GeV and two hadronic taus satisfying  $p_{\textrm{T}}^{\tau} > 25,20$~GeV. 
Based on the kinematic selection used in these ATLAS Run 1 results, it is expected similar selection efficiency for both $Wh$ and $H\to Sh$ 
processes, so these results will be used in the combination. The observed signal strength in each category is determined from a fit to the reconstructed Higgs boson candidate 
mass distribution, resulting in values above unity with relatively large uncertainties, as shown in Table~\ref{tab:AnalysesSummary}. 
  
\begin{figure}[t]
\centering
\includegraphics[width=0.45\textwidth]{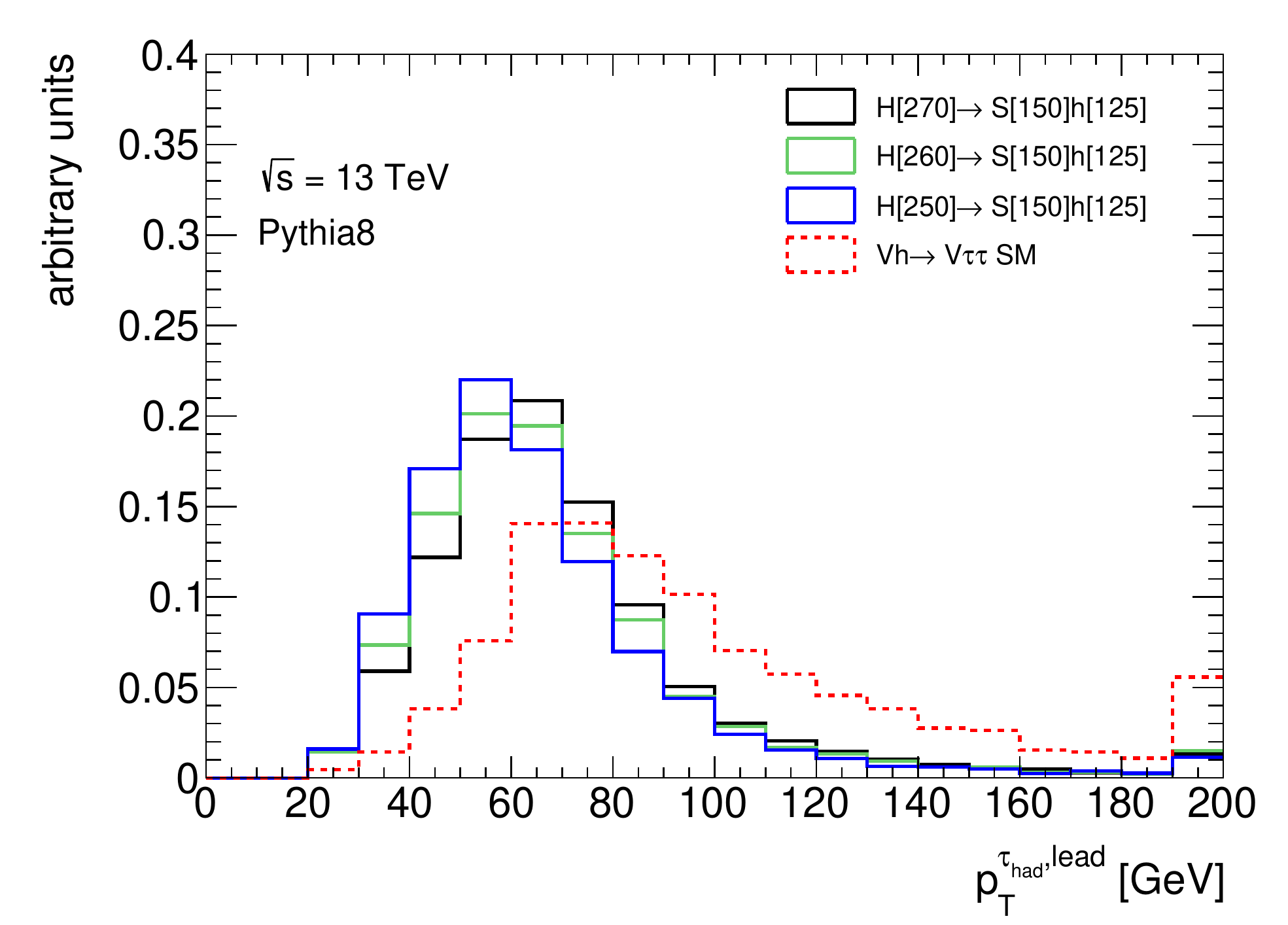}
\caption{Transverse momentum of the leading hadronic tau for several $H\to Sh$ samples (solid lines) compared with the SM $Vh$ 
process with $h\to \tau\tau$ (dashed line) generated with \texttt{PYTHIA8}. The last bin contains overflow events.}
\label{fig:LT_FromTaus}
\end{figure}

Results for the associated production with a $W$ boson of the SM Higgs particle, when it decays to a pair of tau leptons, has been delivered by the CMS experiment using 
Run 1 and Run 2 data~\cite{CMS_Htautau_Run1,CMS_Htautau_Run2}. However, the strategy and event selection is different for each dataset, and in the following they will be 
described. On the one hand, the $\ell\tau_{\textrm{had}}\tau_{\textrm{had}}$ category in CMS Run 1 results makes use of a BDT discriminant based on the $E_{\textrm{T}}^{\textrm{miss}}$ 
and on kinematics related to the di-tau system. In addition, the input discriminating variables include the transverse momentum of the two hadronic taus. Figure~\ref{fig:LT_FromTaus} compares the shapes of the transverse momentum of the leading hadronic tau 
for both $Wh$ and $H\to Sh$ processes. 
Given the fact that the $H \to Sh$ signal tends to be located at the low $p_{\textrm{T}}$ region where the reducible processes such as QCD multilepton, $W/Z$+jes, $W/Z$+$\gamma$, and $t\bar{t}$ mostly contribute, it is expected that the BDT discriminates these backgrounds together with the $H \to Sh$ signal in benefit of the SM $Wh$ process.  
On the other hand, the $e^{\pm}\mu^{\pm}\tau_{\textrm{had}}$ category is further split into two by dividing 
the scalar sum of the leptons' $p_{\textrm{T}}$ at 130~GeV. 
The likelihood fit is performed using the invisible mass of the Higgs decay lepton candidates in each $p_{\textrm{T}}^{e}+p_{\textrm{T}}^{\mu}+p_{\textrm{T}}^{\tau}$ region. 
Figure~\ref{fig:LT_OneHadTau} shows that the contribution for the BSM process concentrates at the low region and the $Wh$ signal is distributed uniformly in these two regions. Due to the fact that the SM backgrounds are dominant in the low region, the statistical fit procedure tends to extract the $Wh$ signal strength from the high region where the $Wh$ signal over background ratio is higher. Since this region has higher impact in the statistical fit it clearly drives the $\mu(Wh)$ result. The BSM hypothesis concentrates at the low region so it is expected that it does not contribute significantly to these results. 
In light of these features, the Run 1 results from CMS for the $Wh$ with $h\to \tau\tau$ are not considered for the signal strength combination in this paper. 

The CMS strategy for the analysis of the Run 2 dataset follows a different approach. The category with one $\tau_{\textrm{had}}$ in the final state requires the scalar 
sum of the $p_{\textrm{T}}$ of the leptons and the $\tau_{\textrm{had}}$ to be above 100~GeV. From Figure~\ref{fig:LT_OneHadTau} it can be seen that the $H \to Sh$ 
efficiency after this cut is applied is above 70\%. The Higgs and $W$ bosons are expected to be close in $\eta$, since they are dominantly produced back-to-back 
in $\phi$ and they may have a longitudinal Lorentz boost. As such, two angular separation cuts between the highest $p_{\textrm{T}}$ lepton and the system formed by 
the $\tau_{\textrm{had}}$ and the remaining lepton are applied. In the $\ell\tau_{\textrm{had}}\tau_{\textrm{had}}$ category, the threshold on the scalar sum of the lepton 
and the two $\tau_{\textrm{had}}$ is 130~GeV. As shown in Figure~\ref{fig:LT_TwoHadTau}, this cut still keeps about 60\% of the $H\to Sh$ process. In addition, the 
vectorial sum of $p_{\textrm{T}}$ of the lepton, the two $\tau_{\textrm{had}}$ candidates and the $E_{\textrm{T}}^{\textrm{miss}}$ is required to be below 70~GeV. Finally, 
only events with small angular separation of the two $\tau_{\textrm{had}}$ candidates in $\eta$ are selected. Given the fact that the event selection is not expected to 
affect the $H\to Sh$ efficiency dramatically, this result should be used in the combination. The observed signal strength for this case presents a deviation with respect to 
the SM expectation of about 1.4$\sigma$, as shown in Table~\ref{tab:AnalysesSummary}.   

\subsection{$Wh\to W\gamma\gamma$}
\label{sec:wyy}
Results for the associated production of a $W/Z$ boson with the SM Higgs particle when the latter  decays into a pair of photons have also been released by the ATLAS 
and CMS collaborations using both Run 1 and Run 2 datasets~\cite{ATLAS_Hyy_Run1,ATLAS_Hyy_Run2_New,CMS_Hyy_Run1,CMS_Hyy_Run2,CMS_Hyy_Run2_FullRun2}. 
The selection criteria in both cases exploit the different vector boson decays by requiring the  presence of leptons, jets or $E_{\textrm{T}}^{\textrm{miss}}$ in the final state. 
The events are classified into three main categories: $Wh$ one-lepton, $Vh$ hadronic and $Vh$ $E_{\textrm{T}}^{\textrm{miss}}$. 

\begin{figure}[t]
\centering
\includegraphics[width=0.45\textwidth]{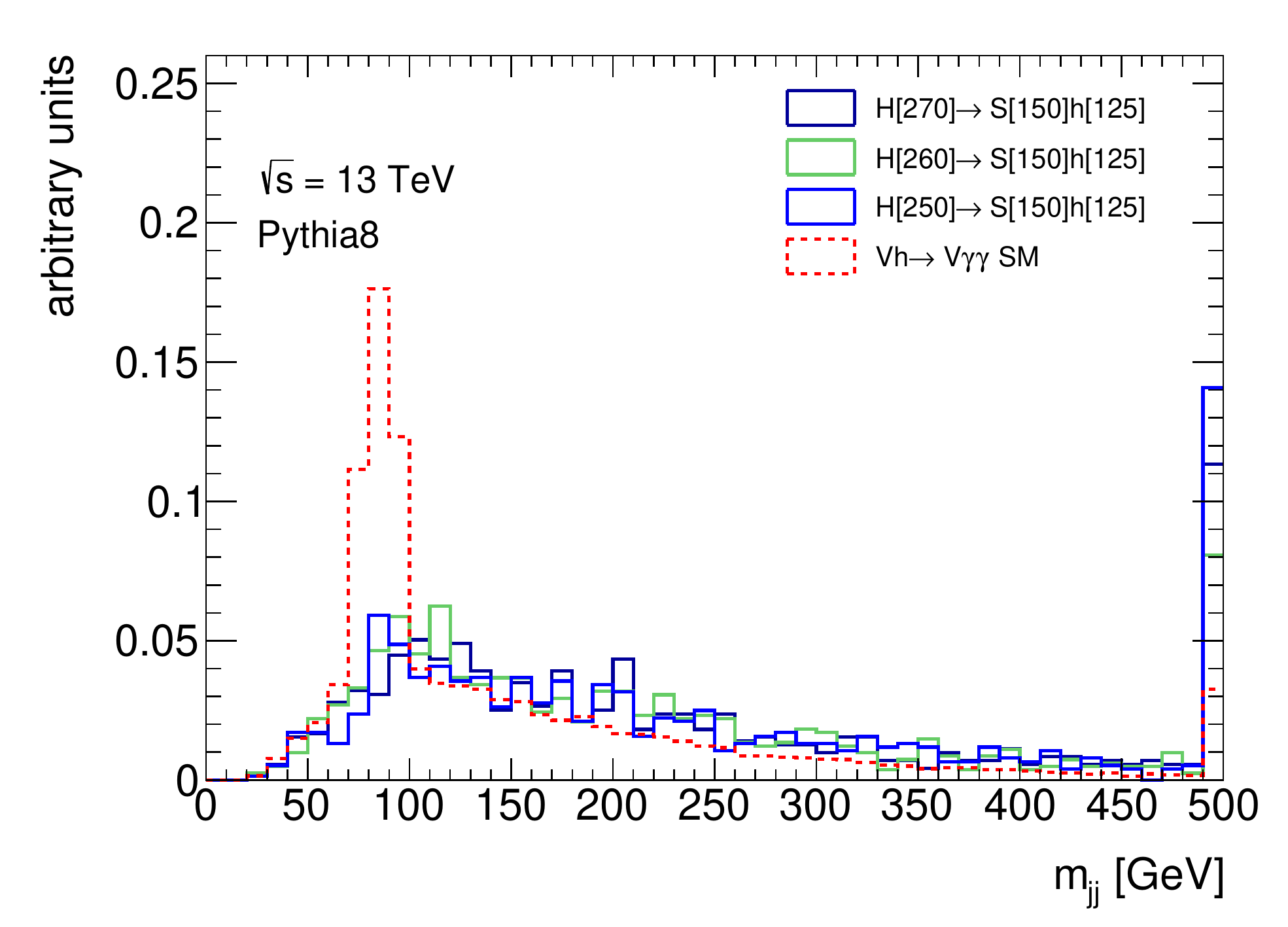}
\caption{Dijet invariant mass in events with two photons and at least two jets for several $H\to Sh$ samples (solid lines) compared with 
the SM $Vh$ process with $h\to \gamma\gamma$ (dashed line) generated with \texttt{PYTHIA8}. The last bin contains overflow events.}
\label{fig:Mjj_yy}
\end{figure}

Events in the $Vh$ hadronic category are required to have a pair of high-energy jets originating from the vector boson decay, hence with $m_{jj}$ consistent with 
the $V$ boson mass. Figure~\ref{fig:Mjj_yy} compares the invariant mass of the dijet system for the SM Higgs boson associated production and the $H\to Sh$ process. 
The selected events contain two photons with $p_{\textrm{T}}^{\gamma_0}>m_{\gamma\gamma}/2$ and $p_{\textrm{T}}^{\gamma_1}>m_{\gamma\gamma}/4$, and at least two jets with transverse momentum above 40~GeV.
For the $Vh$ process the efficiency reaches more than 50\% when selecting an $m_{jj}$ window cut in the range of $[60 - 120]$~GeV. For the BSM process of interest 
here, the $m_{jj}$ selection keeps around 20\% of the total statistics. 

ATLAS Run 1 analysis uses the magnitude of the component of the diphoton momentum transverse to its thrust axis in the transverse plane ($p_{\textrm{Tt}}^{\gamma\gamma}$). 
The strategy selects events with $m_{jj}$ in the $[60 - 110]$~GeV range and $p_{\textrm{Tt}}^{\gamma\gamma}$ above 70~GeV. The ATLAS Run 1 results for the $Vh$ 
hadronic category are not included in the combination due to the high $p_{\textrm{Tt}}^{\gamma\gamma}$ threshold in addition to the restricted  $m_{jj}$ window requirement.   
In Run 2 the ATLAS measurements are carried out using 139~fb$^{-1}$ of $pp$ collision data at $\sqrt{s}= 13$~TeV and the Higgs boson production mechanisms are further 
characterised in terms of the Simplified Template Cross-Section (STXS) framework~\cite{LHC_XSec4,badger2016les,berger2019simplified,amoroso2020les}. In this case two 
$m_{jj}$ regions inclusive in the transverse momentum of the SM Higgs boson are considered. In the first region to tag the hadronic decay of the vector boson the $m_{jj}$ is 
required to be between $[60 - 120]$~GeV, similarly to the Run 1 strategy. This result will not be considered in the combination due to the low acceptance of the BSM process 
in this $m_{jj}$ range, as shown in Figure~\ref{fig:Mjj_yy}. A second STXS region considers events outside the $m_{jj}$ window:  $m_{jj}\, \in \,[0,\,60]\,||\,[120,\,350]$~GeV where 
the majority of the $H\to Sh$ events are expected to contribute. In this case the observed signal strength is $3.16^{+1.84}_{-1.72}$ and this result will be included in the final 
combination. 

\begin{figure}[t]
\centering
\includegraphics[width=0.45\textwidth]{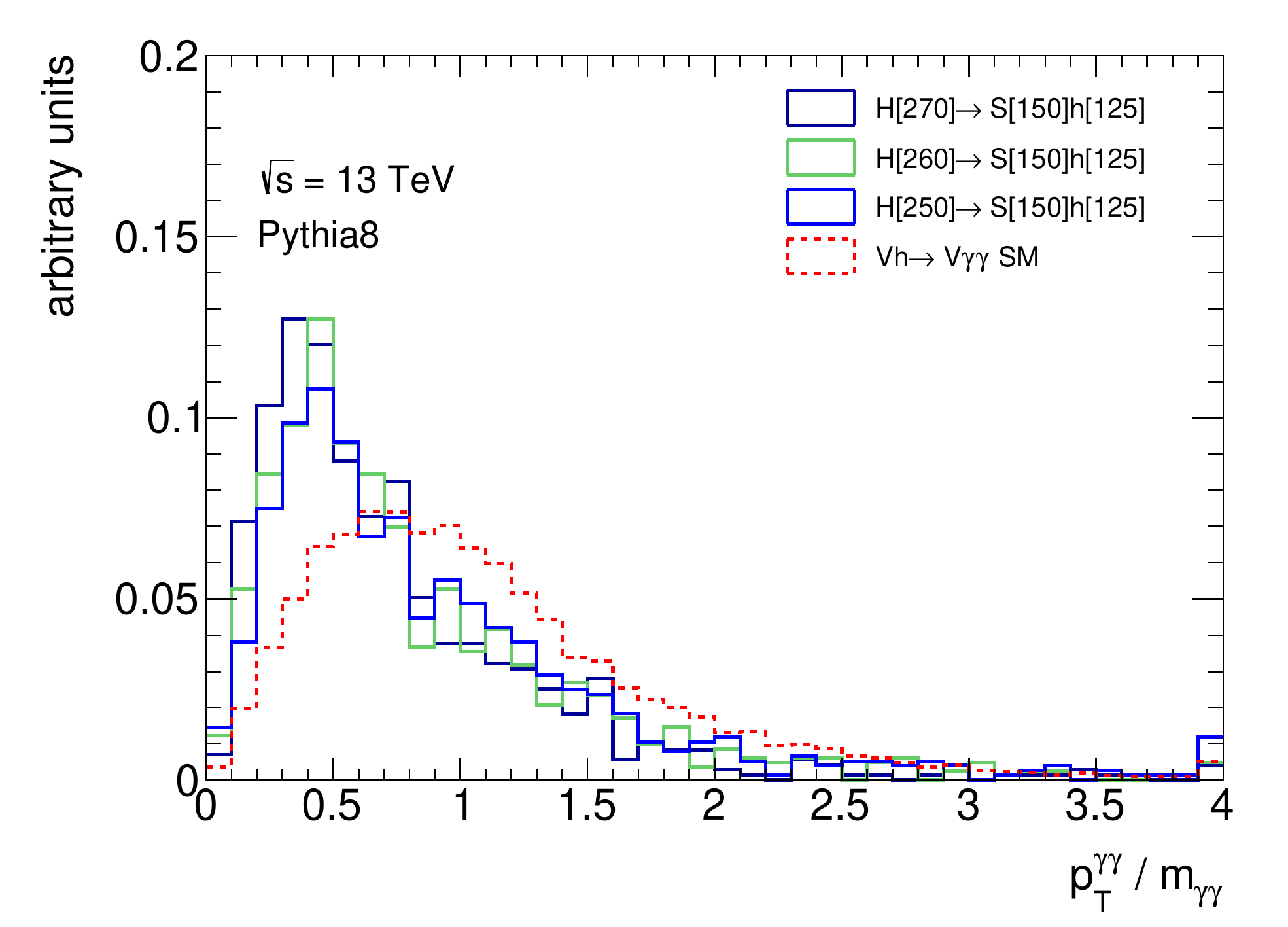}
\caption{Ratio between the transverse momentum and the invariant mass of the diphoton system in events with at least two jets for several $H\to Sh$ 
samples (solid lines) compared with the SM $Vh$ process with $h\to \gamma\gamma$ (dashed line) generated with \texttt{PYTHIA8}. 
The last bin contains overflow events.}
\label{fig:yy_Ptyy_Myy}
\end{figure}

Results from CMS make use of the angle between the diphoton and the diphoton-dijet system in 
both Run 1 and Run 2 datasets. The main difference between the CMS strategies is the use of the $p_{\textrm{T}}^{\gamma\gamma}/m_{\gamma\gamma}$ quantity.
In CMS Run 1~\cite{CMS_Hyy_Run1} analysis, events are required to satisfy $p_{\textrm{T}}^{\gamma\gamma} > 13m_{\gamma\gamma}/12$ for the $Vh$ hadronic 
category. Figure~\ref{fig:yy_Ptyy_Myy} shows the ratio between the diphoton transverse momentum and its invariant mass. The $p_{\textrm{T}}^{\gamma\gamma}/m_{\gamma\gamma}$
requirement highly reduces the $H\to Sh$ acceptance by rejecting more than 85\% of the BSM events. Similarly, the full Run 2 strategy~\cite{CMS_Hyy_Run2_FullRun2} considers the
$p_{\textrm{T}}^{\gamma\gamma}/m_{\gamma\gamma}$ quantity as input variable in the BDT. Due to the SM $Vh$ spectrum in Figure~\ref{fig:yy_Ptyy_Myy} it is expected that the BDT
discriminates the low $p_{\textrm{T}}^{\gamma\gamma}/m_{\gamma\gamma}$ region where the background and the $H\to Sh$ processes dominate. In light of this, the CMS 
Run 1 as well as the full Run 2   results will not be considered in the combination. However, the $p_{\textrm{T}}^{\gamma\gamma}/m_{\gamma\gamma}$  requirement was dropped 
in the partial Run 2 results using 35.9~fb$^{-1}$~\cite{CMS_Hyy_Run2}. The measurement for the $Vh$ hadronic category in this case presents a deviation from the SM expectation 
of approximately 1.5$\sigma$, being the observed signal strength $5.1^{+2.5}_{-2.3}$. This result will be included in the final combination. 

The $Vh$ $E_{\textrm{T}}^{\textrm{miss}}$ category is enriched in events with a leptonic decay of the $W$ boson, when the lepton is not detected or does not satisfy the 
selection criteria (denoted by  $\cancel{\ell}$), or with a $Z$ boson decaying into a pair of neutrinos. In this case the selection criteria relies on the $E_{\textrm{T}}^{\textrm{miss}}$ 
distribution to select events in the high range. The strategy from CMS uses a lower bound of 85(70)~GeV on the $E_{\textrm{T}}^{\textrm{miss}}$ for Run 2(1) results. 
Similarly, ATLAS Run 1 results are obtained by applying a cut on a  $E_{\textrm{T}}^{\textrm{miss}}$ based quantity which is approximately  equivalent to a 
$E_{\textrm{T}}^{\textrm{miss}} > 70-100$~GeV requirement. 

\begin{figure}[t]
\centering
\includegraphics[width=0.45\textwidth]{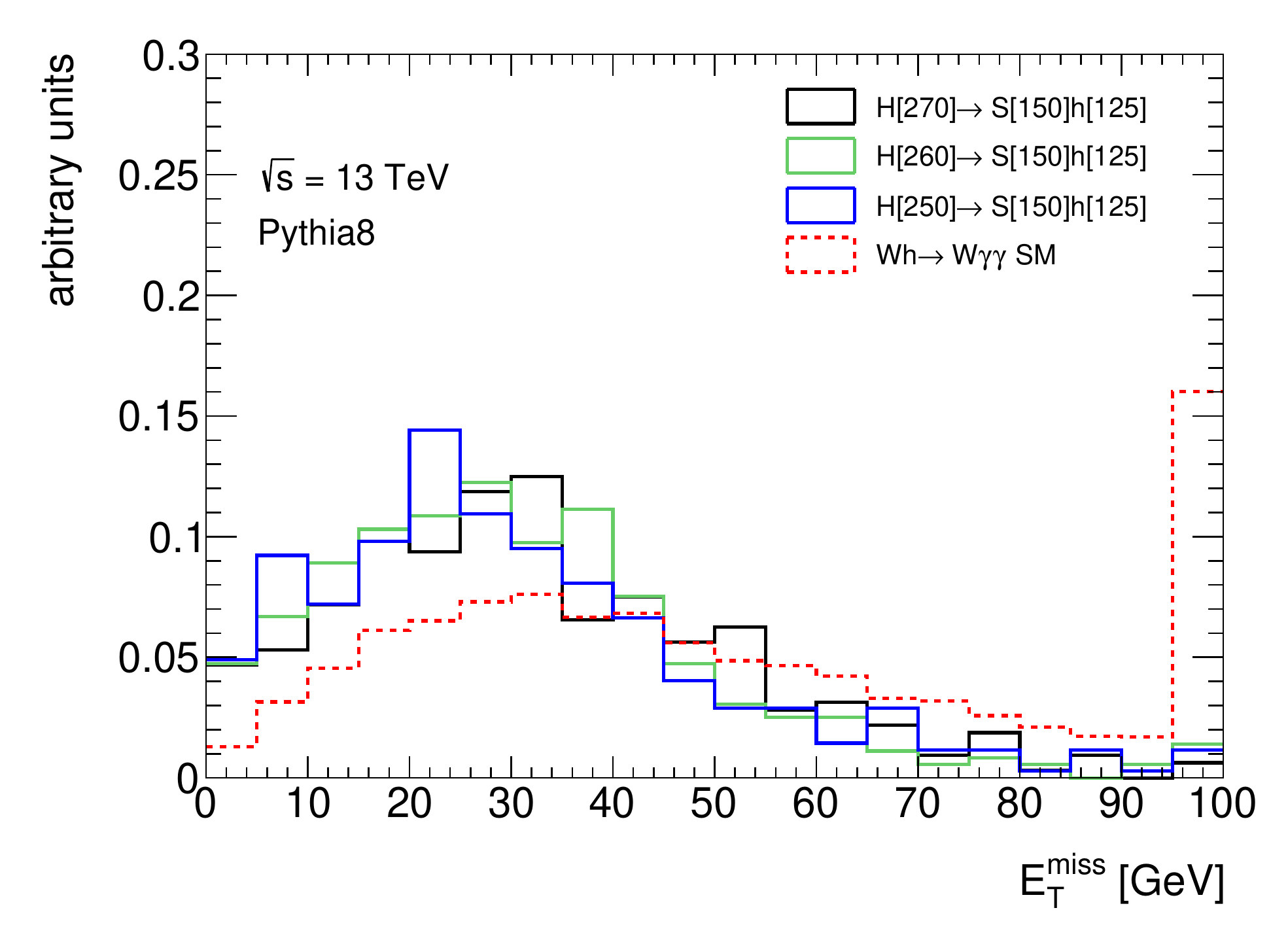}
\caption{Missing transverse energy in events with two photons and one electron or muon for several $H\to Sh$ samples (solid lines) compared 
with the SM $Wh$ process with $h\to \gamma\gamma$ (dashed line) generated with \texttt{PYTHIA8}. The last bin contains overflow events.}
\label{fig:yy_MET_WHLep}
\end{figure}

\begin{figure}[t]
\centering
\subfloat[]{\includegraphics[width=0.45\textwidth]{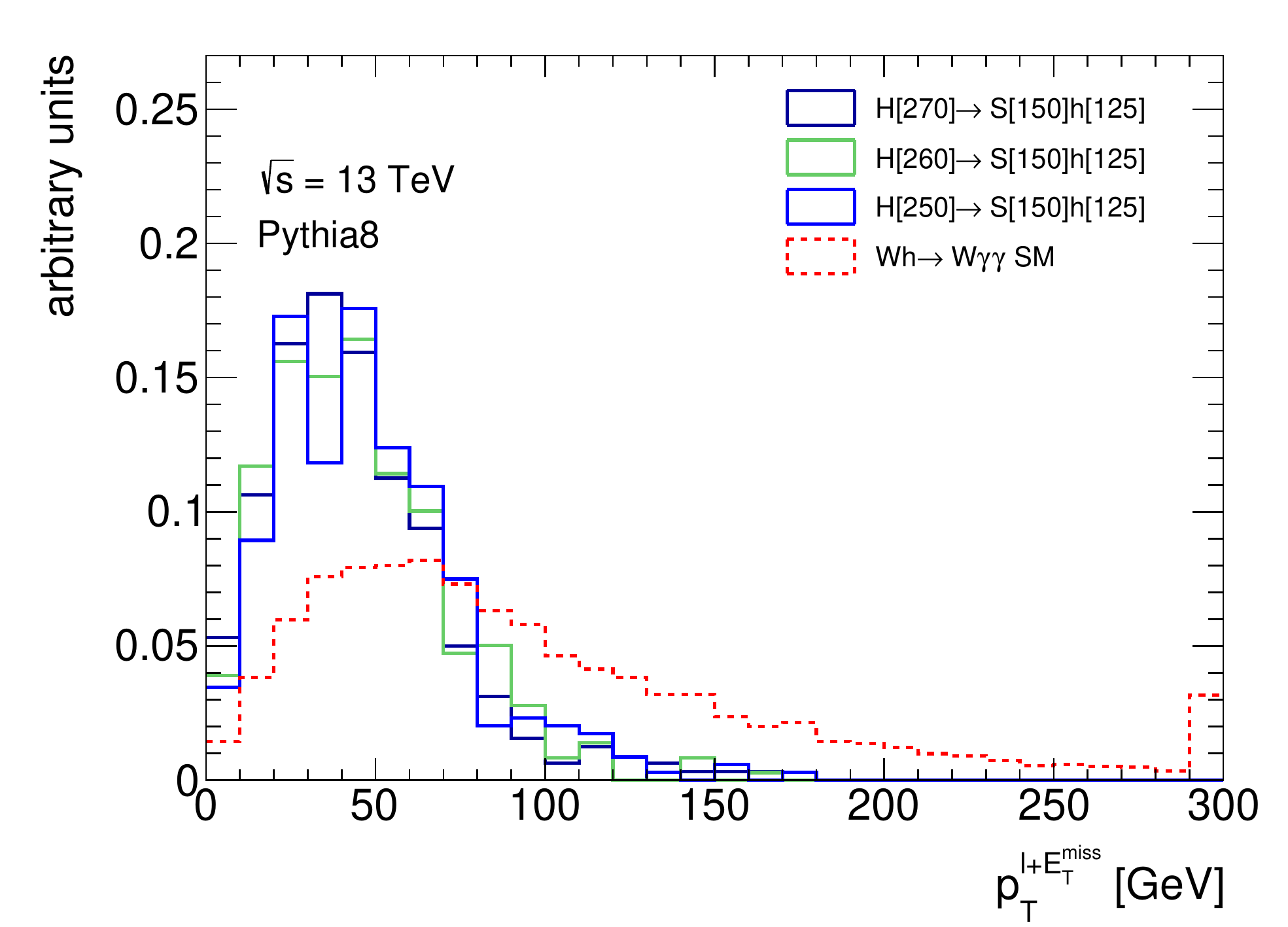}
\label{fig:PtlMET}}\\
\subfloat[]{\includegraphics[width=0.45\textwidth]{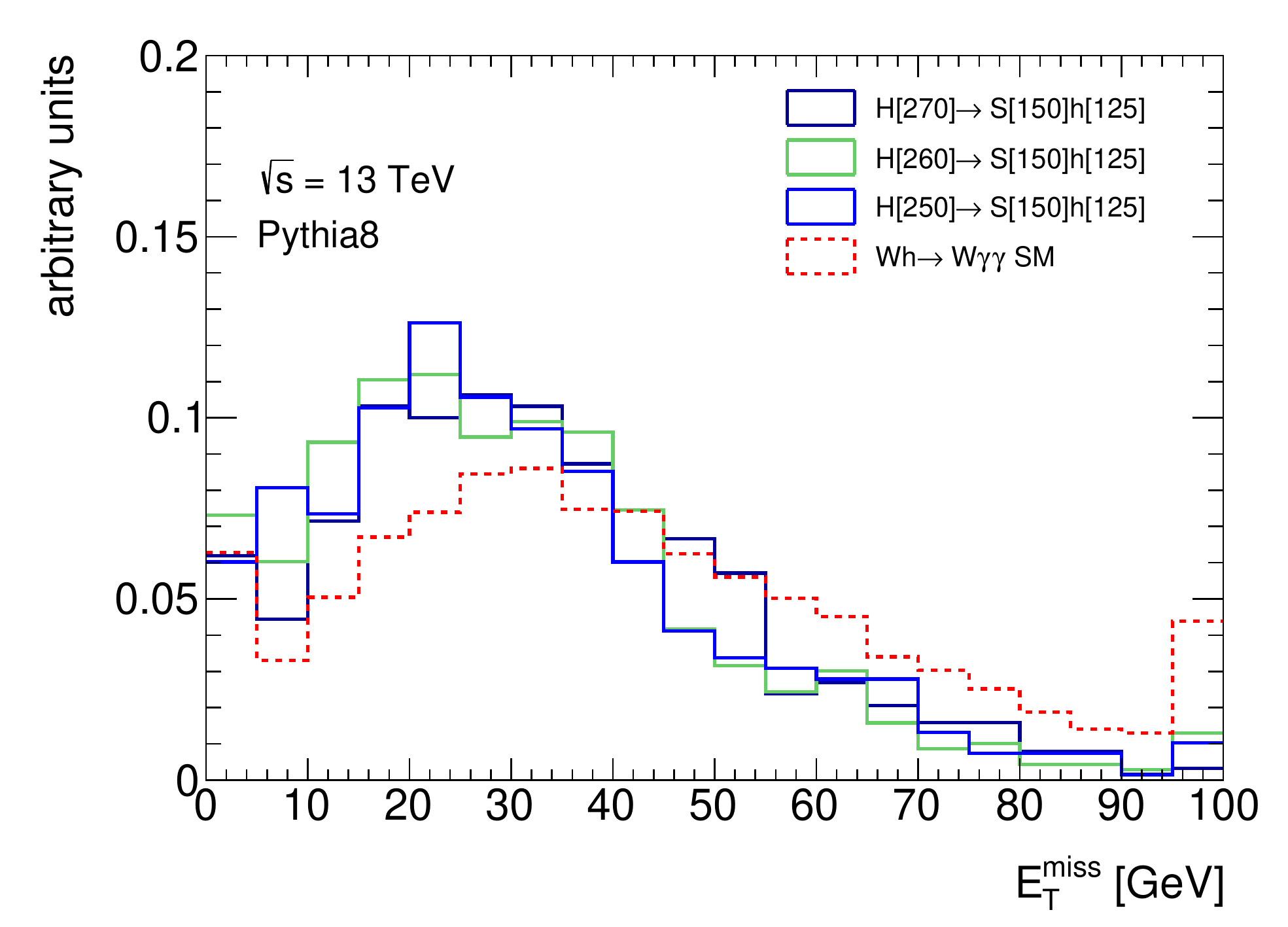}
\label{fig:yy_MET_WHLep_PtW150}}
\caption{Transverse momentum of the lepton and the $E_{\textrm{T}}^{\textrm{miss}}$ system in events with two photons and a lepton (a) and 
missing transverse energy  after requiring $p_{\textrm{T}}^{\ell + E_{\textrm{T}}^{\textrm{miss}}} < 150$~GeV (b) for several $H\to Sh$ samples 
(solid lines) compared with the SM $Wh$ process with $h\to \gamma\gamma$ (dashed line) generated with \texttt{PYTHIA8}. The last bin 
contains overflow events.}
\end{figure}

The $Wh$ one-lepton class is characterised by a leptonically decaying $W$ boson, hence it targets events with two photons accompanied by one electron or one muon. 
CMS further splits the one-lepton category by dividing the  $E_{\textrm{T}}^{\textrm{miss}}$ spectrum at 45~GeV~\cite{CMS_Hyy_Run1}. Figure~\ref{fig:yy_MET_WHLep} shows the missing 
transverse energy for events with two photons and a lepton. At this cut value, the $Wh$ process is divided by 50\% in each region, with the events in the high 
$E_{\textrm{T}}^{\textrm{miss}}$ range the ones driving the result on the measured signal strength. The $H\to Sh$ signal acceptance is approximately 20\% in the 
high $E_{\textrm{T}}^{\textrm{miss}}$  region. The CMS Run 1 results will be discarded in the combination as they are computed including not only the $Vh$ one-lepton 
category but also the hadronic and $E_{\textrm{T}}^{\textrm{miss}}$ ones as well. Conversely, CMS full Run 2 results are produced in the Higgs STXS framework and delivered 
for the one-lepton and the hadronic categories separately. In addition, two regions are defined using the transverse momentum of the $V$ boson ($p_{\textrm{T}}^{l+E_{T}^{miss}}$) 
at 75~GeV for the leptonic category. Only the  $p_{\textrm{T}}^{l+E_{T}^{miss}}<75$~GeV result will be considered as the measured signal strengths are provided for each 
analysis category and the contribution of the BSM process is dominant in the low region, as shown in Figure~\ref{fig:PtlMET}. The Run 2 CMS result measures an observed 
signal strength for the $Wh$ one-lepton category of  $1.31^{+1.42}_{-1.12}$~\cite{CMS_Hyy_Run2_FullRun2}. 

The full Run 2 ATLAS strategy for the $Wh$ leptonic category builds a BDT with photon and lepton variables used as input~\cite{ATLAS_Hyy_Run2_New}. In addition, $E_{\textrm{T}}^{\textrm{miss}}$ 
related quantities and vector-boson kinematics are also used as input variables in the BDT. The  $Wh$ one-lepton events  are split using the transverse momentum 
of the lepton and the $E_{\textrm{T}}^{\textrm{miss}}$ at 150~GeV. Figure~\ref{fig:PtlMET} compares the shape of the $p_{\textrm{T}}^{l+E_{T}^{miss}}$ quantity 
for $Wh$ and $H\to Sh$ processes. The contribution of the BSM signal in the high region of the distribution is expected to be negligible so this result will not be considered 
in the combination. However the $H\to Sh$ process is almost entirely located at $p_{\textrm{T}}^{l+E_{T}^{miss}} < 150$~GeV. Since the $E_{\textrm{T}}^{\textrm{miss}}$ is 
used in the BDT it is important to verify that in the low  $p_{\textrm{T}}^{l+E_{T}^{miss}}$ region the performance of the distribution is similar for the $H\to Sh$ and 
SM $Wh$ processes. Figure~\ref{fig:yy_MET_WHLep_PtW150} shows the $E_{\textrm{T}}^{\textrm{miss}}$ distribution in events with two photons and one electron or 
muon after requiring $p_{\textrm{T}}^{\ell + E_{\textrm{T}}^{\textrm{miss}}} < 150$~GeV. It can be observed that the spectrum for each process is similar being the mean 
of the distributions 39~GeV and 31~GeV for the SM $Wh$ and $H\to Sh$ signals, respectively. The full Run 2 ATLAS result in the low $p_{\textrm{T}}^{l+E_{T}^{miss}}$ 
phase space presents a deviation from the SM value of $\sim 2\sigma$. The observed signal strength is $2.41^{+0.71}_{-0.70}$ and this measurement will be included in 
the combination. The ATLAS strategy for the Run 1 dataset selects $Wh$ one-lepton events by applying a cut on a $E_{\textrm{T}}^{\textrm{miss}}$ related quantity.  
In light of this requirement and the difference between the SM and BSM processes as shown in Figure~\ref{fig:yy_MET_WHLep}, the ATLAS Run 1 results for the one-lepton 
category are not included in the final combination. 

\subsection{$Wh\to WZZ$}
\label{sec:wzz}
\begin{figure}[t]
\centering
\includegraphics[width=0.45\textwidth]{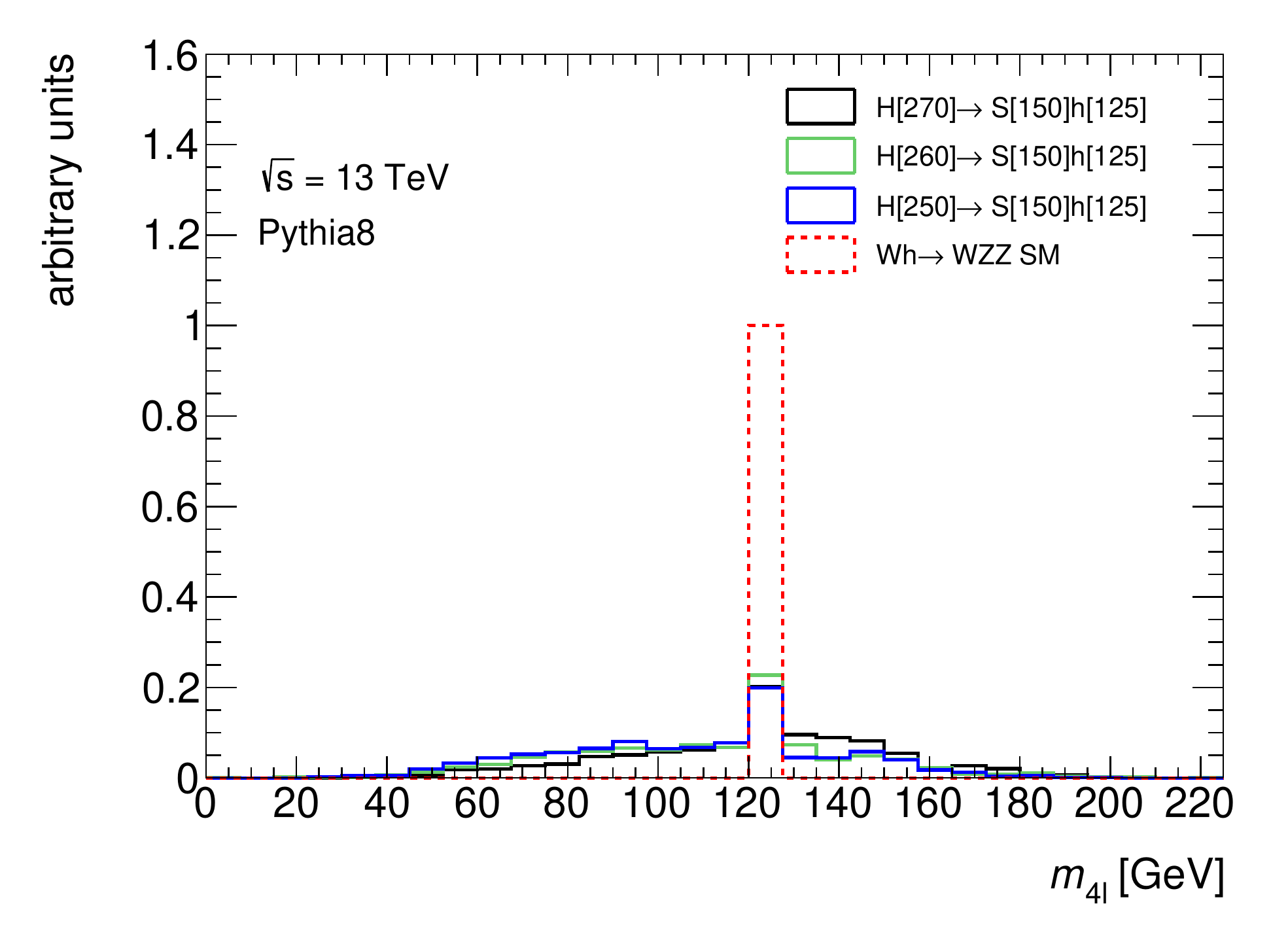}
\caption{Invariant mass of the four leptons for several $H\to Sh$ samples (solid lines) compared with the SM $Wh$ process 
with $h\to ZZ \to 4\ell$ (dashed line) generated with \texttt{PYTHIA8}. The last bin contains overflow events. }
\label{fig:M4l}
\end{figure}

\begin{figure}[t]
\centering
\subfloat[]{\includegraphics[width=0.45\textwidth]{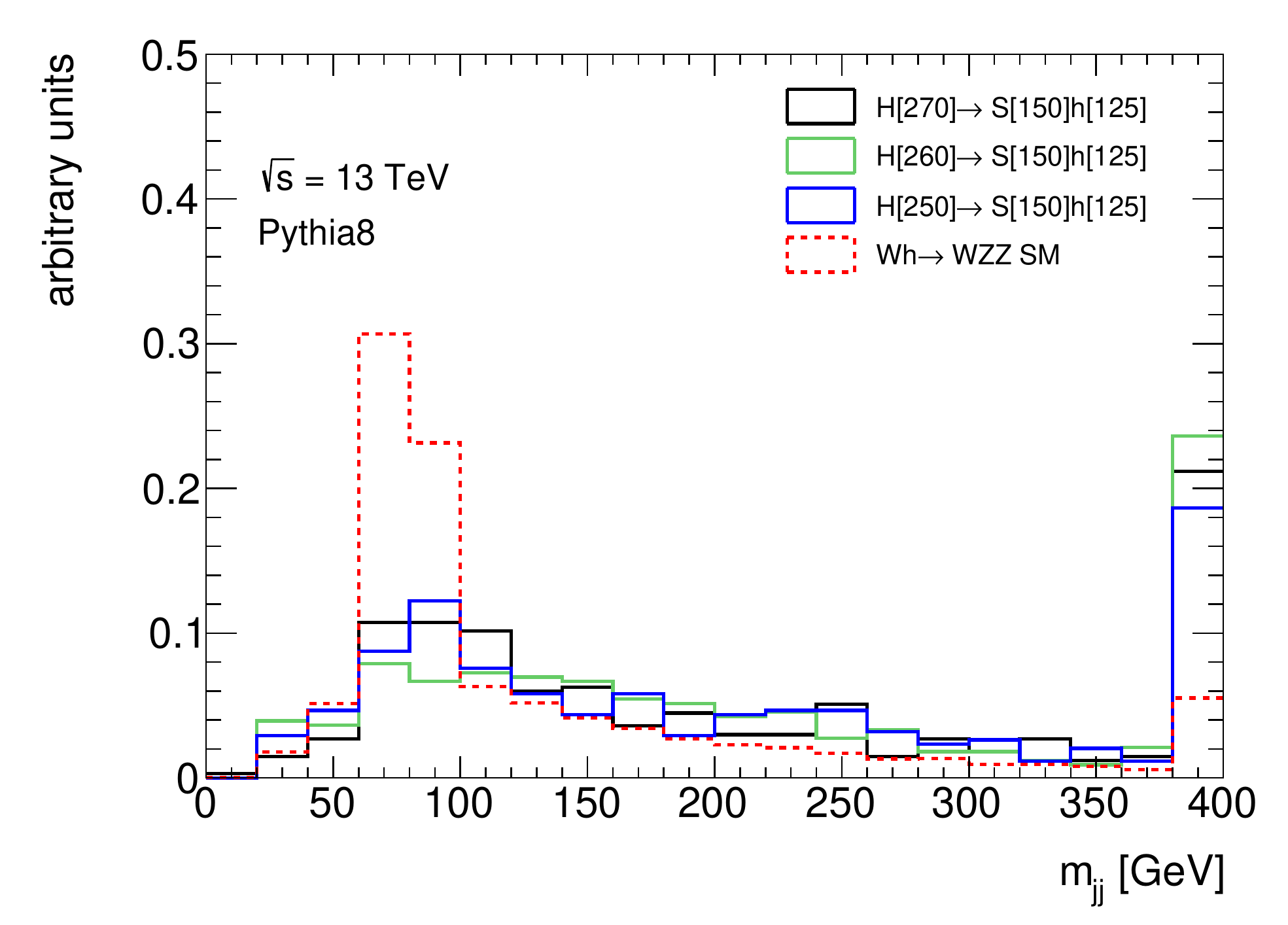}
\label{fig:Mjj}}\\
\subfloat[]{\includegraphics[width=0.45\textwidth]{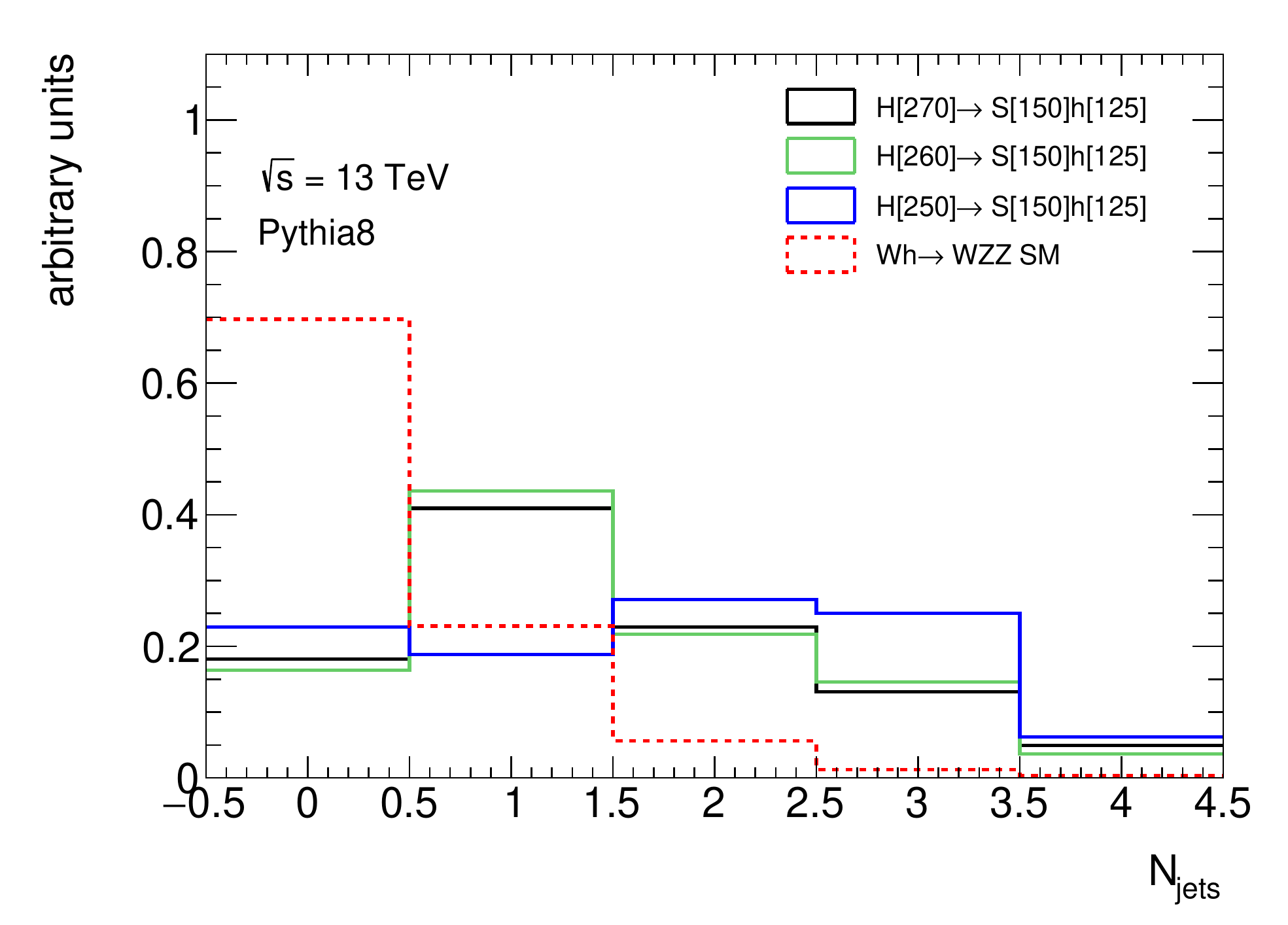}
\label{fig:Njets_HZZ}}
\caption{Dijet invariant mass (a) and jet multiplicity (b) for several $H\to Sh$ samples (solid lines) compared with the SM $Wh$ process 
with $h\to ZZ \to 4\ell$ and $W\to q\bar{q}$ (dashed line) generated with \texttt{PYTHIA8}. The last bin contains overflow events. }
\end{figure}

ATLAS and CMS results for the $H\to ZZ^{\ast} \to 4\ell$ decay mode using the full Run 2 dataset are published in Ref.~\cite{ATLAS_ZZ_Run2} and Ref.~\cite{CMS_ZZ_Run2}, 
respectively. The common strategy makes use of the invariant mass of the four leptons from the Higgs decay ($m_{4\ell}$) to select the Higgs candidates in a 
window around its mass: 115~GeV$< m_{4\ell} < 130$~GeV.  Approximately 70\% of the $H\to Sh$ events are outside this $m_{4\ell}$ mass window so  this requirement 
highly reduces the acceptance of the BSM signal as shown in Figure~\ref{fig:M4l}. Both experiments split the events depending on the hadronic or leptonic decay of the $V$ 
boson produced in association with the Higgs boson. In the hadronic channel, the four leptons from the Higgs decay are accompanied by two jets and the $m_{jj}$ distribution 
is exploited. CMS selects events in the window around the $W$/$Z$ mass peak: 60~GeV$<m_{jj}<120$~GeV and ATLAS uses the $m_{jj}$ spectrum as input in a neural 
network (NN) to separate between the $Vh$ and VBF production mechanisms. Given the dependence of the SM results on the $m_{jj}$ spectrum it is expected that these 
measurements do not include the $H\to Sh$ signal. Figure~\ref{fig:Mjj} compares the $m_{jj}$ distribution for the SM $Wh$ and the $H\to Sh$ processes for events with 
four leptons and two jets in the final state. The rejection of the BSM process is approximately 70\% when requiring events within the range 60~GeV$<m_{jj}<120$~GeV. 
Since ATLAS and CMS strategies rely on the $m_{jj}$ window the results for the hadronic category will not be included in the combination. 

In the $Wh$ leptonic category, the analyses require an extra lepton in the final state.  ATLAS strategy uses variables as the jet and $b$-tagged jet multiplicities, 
in addition to the $E_{\textrm{T}}^{\textrm{miss}}$ distribution, to build a MVA discriminant to distinguish between $Vh$ and $tth$ production mechanisms. 
Figure~\ref{fig:Njets_HZZ} compares the distributions of the expected number of jets with $p_{\textrm{T}} > 30$~GeV for the SM and the BSM processes from MC simulation. 
Events from the $Wh$ decay are dominant at low jet multiplicities, being the contribution of events with zero jets of around 70\%. Conversely, the $H\to Sh$ signal 
tends to have higher number of jets and it only contributes $\sim$20\% in events with no jets in the final state. Due to the expected differences in the jet multiplicity 
distribution between the SM and BSM processes the ATLAS results for the leptonic category are not combined with the rest of $Wh$ results.  

\begin{figure}[t]
\centering
\includegraphics[width=0.45\textwidth]{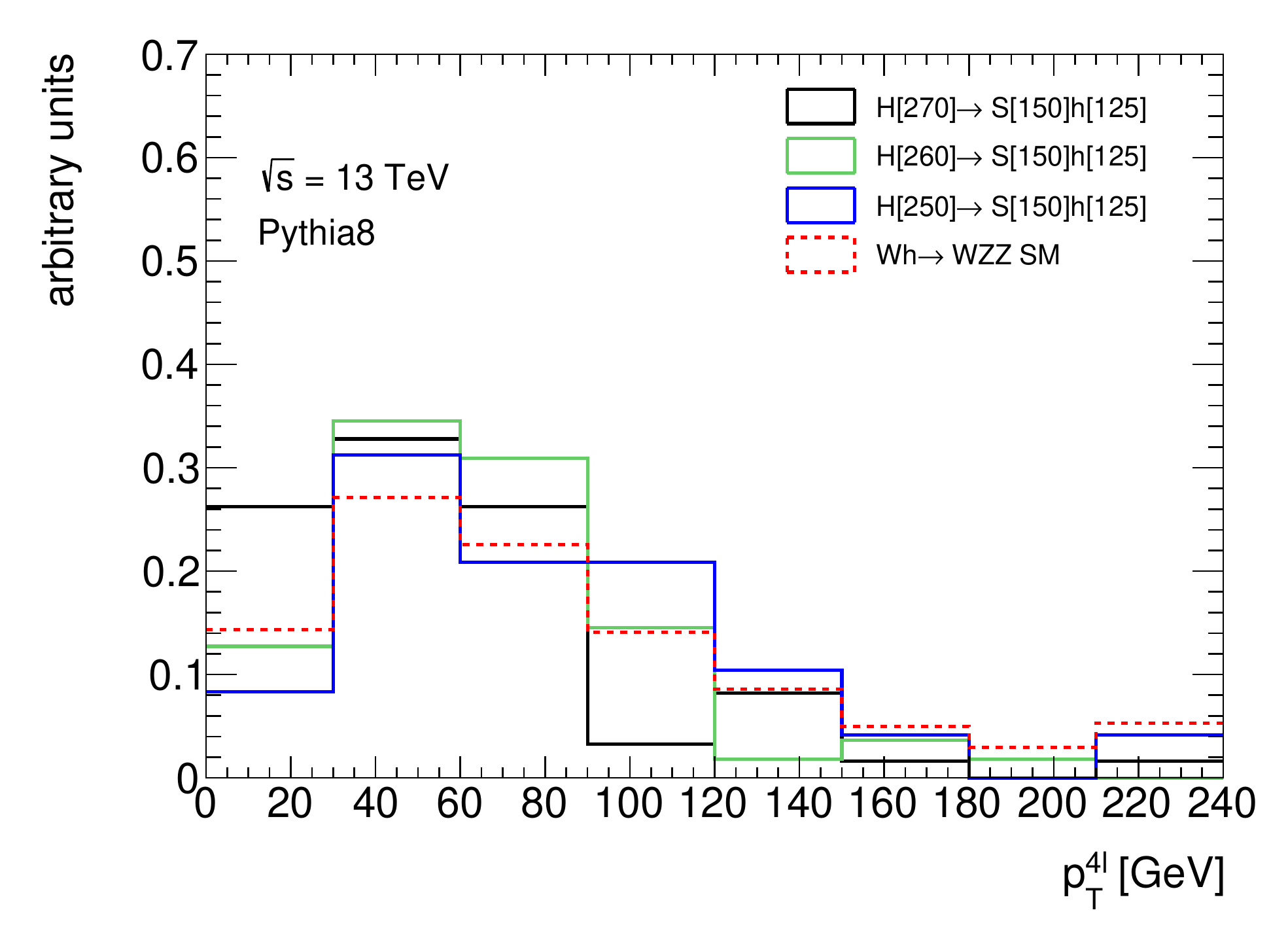}
\caption{Transverse momentum of the leptons associated to the SM Higgs decay for the $H\to Sh$ samples (solid lines) compared with the SM $Wh$ 
process with $h\to ZZ \to 4\ell$ and $W\to \ell\nu$ (dashed line) generated with \texttt{PYTHIA8}. The last bin contains overflow events. }
\label{fig:pT4L_HZZ}
\end{figure}

Finally, for the leptonic category CMS selects events with at most three jets, hence it is expected a high acceptance of the $H\to Sh$ signal which can be seen from  Figure~\ref{fig:Njets_HZZ}. 
In addition, the final candidate events are split into two regions of the Higgs transverse momentum: $p_{\textrm{T}}^{4\ell} < 150$~GeV and $p_{\textrm{T}}^{4\ell} > 150$~GeV. 
Figure~\ref{fig:pT4L_HZZ} shows the transverse momentum of the four leptons associated to the SM Higgs decay. For both SM $Wh$ and $H\to Sh$ processes the bulk of 
the events is located in the low $p_{\textrm{T}}^{4\ell}$ region. 
In light of the  $p_{\textrm{T}}^{4\ell}$ distribution, 
only the measured signal strength for the $p_{\textrm{T}}^{4\ell} < 150$~GeV region will be included in the final combination. The observed cross section in this case normalised 
to the SM prediction results in $3.21^{+2.49}_{-1.85}$ from Ref.~\cite{CMS_ZZ_Run2}. 

\section{Compatibility with inclusive observables}
\label{sec:inclusive}

While this paper focuses on the anomalous production of the SM Higgs boson in association with leptons, it is relevant to investigate if these findings do not 
contradict measurement of inclusive observables made by the experiments. It is known that the additional production of the SM Higgs boson via the $H \to Sh$
process would distort the $h$ transverse momentum and the rapidity spectra. The transverse momentum would be enhanced  at moderate values. The SM 
Higgs boson would be produced more centrally. A survey of available Run 1 and Run 2 data was performed~\cite{ATLAS_HZZ4l_run2,ATLAS_HZZ4l_run1,ATLAS_Hyy_run2,CMS_HZZ4l_run2,CMS_Hyy_HZZ_combined_run1,CMS_Hyy_run1,CMS_HWW_run2}. 
All data sets, except the ATLAS Run 2 $h\to ZZ^{*}\to 4\ell$ results, display compatibility with these features concurrently. While the overall deviation from the SM hypothesis 
is not statistically significant (of order of two standard deviations), it is compatible with the hypothesis considered here. A comprehensive analysis of inclusive observables 
will be performed when the complete Run 2 data set is available. 

\section{Results and Conclusions}
\label{sec:results}

The interpretation of the multi-lepton anomalies at the LHC reported in Refs.~\cite{vonBuddenbrock:2017gvy,vonBuddenbrock:2019ajh} with the decay $H\to Sh$ 
predicts anomalously large values of the signal strength of $Wh$. This effect should be visible with the available results from ATLAS and CMS so far. Section~\ref{sec:method} 
provides a comprehensive synopsis of the current status of the search and measurements of $Wh$ production in the SM, where the available results correspond to the Run 1 and, 
partial or complete, Run 2 data sets. Table~\ref{tab:AnalysesSummary} gives the summary of the available results and indicates which ones are used in the combination 
with the appropriate explanation. The combination is estimated as the error weighted signal strength of each considered result. The uncertainties between different channels, for both experiments and across data sets are treated as uncorrelated. 
The obtained result is then compared with the one expected for the SM scenario with $\mu = 1$.  
By using the method given in Particle Data Group to combine different measurements with asymmetric uncertainties~\cite{pdg2020} the combined $Wh$ signal strength from Table~\ref{tab:AnalysesSummary} results in  $\mu(Wh)_{Inc} = 2.41 \pm 0.37$ which corresponds to a deviation from the SM of 3.8$\sigma$. 
\footnote{Also by removing the channels that are based on BDTs, such as in Ref.~\cite{ATLAS_HWW_Run2} and the one-lepton category in Ref.~\cite{ATLAS_Hyy_Run2_New}, the combined result is $\mu(Wh)_{\rm No-BDT} = 2.39 \pm 0.44$ which corresponds to a deviation of 3.2$\sigma$ from the SM.}
The errors are dominated by statistical and experimental uncertainties, which are uncorrelated. The bulk of the correlated uncertainties pertain to the theoretical error, which for this production mechanism is significantly smaller than the error claimed here.

As discussed in Section~\ref{sec:method}, the estimate made here is based on searches and measurements biased towards the SM. 
The combination of the rejected measurements from Table~\ref{tab:AnalysesSummary} results in  $\mu(Wh)_{Rej}=0.95\pm0.35$. 
In the corners of the phase-space where the BSM signal is not expected to contribute, the signal strength of the $Wh$ production is consistent with the SM prediction. 
Combining all the results provides a signal strength of $\mu(Wh)_{All}=1.64\pm0.25$, which corresponds to a deviation from the SM value of unity of $2.6\sigma$. 

The impact on the measurement of $h$ cross-sections due to the BSM signal considered here goes beyond the associated production of leptons, as discussed here. 
The measurement of the Higgs boson transverse momentum and rapidity will also be affected. These effects will be studied with results with the full Run 2 data set, when available. 
While the effect seen here seems in qualitative agreement with the multi-lepton anomalies interpreted with the simplified model described in Section~\ref{sec:model}, it is 
important to confront the value of $\mu(Wh)_{Inc}$ with that expected with the ansatz of Br$(H\to Sh)=100\%$ made in Refs.~\cite{vonBuddenbrock:2017gvy,vonBuddenbrock:2019ajh}. 
Assuming the cross-section $\sigma(H\to S^*h)=10$\,pb~\cite{Fang:2017tmh}, where $h$ is on-shell, one would expect a combined (including the SM) signal strength of about 6 for the combination 
of the channels considered in Section~\ref{sec:method}. This is considerably larger than the signal strength observed here, notwithstanding the expected bias discussed in
Section~\ref{sec:method}. This indicates that explaining the multi-lepton anomalies reported in Refs.~\cite{vonBuddenbrock:2017gvy,vonBuddenbrock:2019ajh} would require 
a considerable contribution from $H\to SS$ along with $H\to Sh$. The decay $H\to hh$ would be suppressed due to results from direct searches. 

Irrespective of the size of $\mu(Wh)_{Inc}$ determined here, one needs to seriously consider a situation whereby the production of $h$ at the LHC is contaminated with 
production mechanisms other than those predicted in the SM. This implies that the determination of couplings of $h$ to SM particles would be seriously compromised by 
model dependencies. This further enhances the physics case of Higgs factories on the basis of $e^+e^-$~\cite{Baer:2013cma,CEPCStudyGroup:2018ghi,Abada:2019zxq} 
and $e^-p$~\cite{AbelleiraFernandez:2012cc,Han:2009pe,Biswal:2012mp,Agostini:2020fmq} collisions, while the potential for the direct observation of new physics at 
the HL-LHC is enriched strongly. The production of $H$ in $e^-p$ collisions would be suppressed, therefore, the  determination of the Higgs boson couplings would be less 
model dependent compared to proton-proton collisions. Assuming the current value of the $h$ global signal strength at the LHC, and that the couplings of $h$ to SM particles 
are as in the SM, the contamination at the LHeC would be five times smaller than that at the LHC~\cite{Mosomane:2017jcg}.  The LHeC, with input from proton-proton collisions, 
would allow for the precise determination of the $hWW$ coupling, which combined with the superb measurement of the $hZZ$ coupling in $e^+e^-$ collisions, would provide a
 powerful probe into EWSB. 

\section{Acknowledgements}
The authors are grateful for support from the South African Department of Science and Innovation through the SA-CERN program and the National Research 
Foundation for various forms of support. The authors are also indebted to the Research Office of the University of the Witwatersrand for grant support. This 
work was supported by the Beijing Municipal Science and Technology Commission project (Grant No.: Z191100007219010).


\end{document}